\documentclass[a4paper, reqno]{amsart}

\usepackage[utf8]{inputenc}	
\usepackage[T1]{fontenc}	
\usepackage{a4wide}
\usepackage[dvipsnames]{xcolor}

\usepackage{amsmath}		
\usepackage{amssymb}		
\usepackage{enumerate}
\usepackage{nicefrac}	
\usepackage{array}

\usepackage{booktabs}
\usepackage{calc}			
\usepackage{flafter}		
\usepackage{esdiff}
\usepackage{pgffor}  
\usepackage{pgf}
\usepackage{mathabx}
\usepackage{multicol}
\usepackage{multirow}
\usepackage{longtable}
\usepackage{appendix}

\usepackage{changes} 
\definechangesauthor[name={David Alonso},       color=ballblue ]{da}
\definechangesauthor[name={Steffen Bauer},      color=PineGreen]{sb}
\definechangesauthor[name={Markus Kirkilionis}, color=RoyalBlue]{mk}
\definechangesauthor[name={Lisa Maria Kreusser},color=red      ]{lk}
\definechangesauthor[name={Luca Sbano},         color=purple   ]{ls}
\colorlet{Changes@Color}{blue}

\usepackage{placeins} 
\makeatletter
\AtBeginDocument{%
  \expandafter\renewcommand\expandafter\subsection\expandafter
    {\expandafter\@fb@secFB\subsection}%
  \newcommand\@fb@secFB{\FloatBarrier
    \gdef\@fb@afterHHook{\@fb@topbarrier \gdef\@fb@afterHHook{}}}%
  \g@addto@macro\@afterheading{\@fb@afterHHook}%
  \gdef\@fb@afterHHook{}%
}
\makeatother

\usepackage{hyperref}
\usepackage{hyperxmp}
\usepackage{textcomp}	
\usepackage{amscd}	
\usepackage{amsaddr}			
\usepackage{mathrsfs}		
\usepackage{graphicx}

\usepackage[subrefformat=simple,labelformat=simple]{subcaption}
\newcommand*\ccol[1]{\omit\hfil$\displaystyle#1$\hfil\ignorespaces} 

\newcommand\N{\mathbb{N}}
\newcommand\R{\mathbb{R}}

\usepackage{bbm}

\newcommand*\di{\mathop{}\!\mathrm{d}}

\renewcommand\epsilon{\varepsilon}

\renewcommand\theta{\vartheta}

\newcommand*{\num}[1]{{#1}}
\newcommand*{\con}[1]{{\mathrm{#1}}}

\newcommand*{\nS}{\num{S}}
\newcommand*{\nI}{\num{I}}
\newcommand*{\nR}{\num{R}}
\newcommand*{\nD}{\num{D}}

\newcommand*{\cS}{\con{S}}
\newcommand*{\cI}{\con{I}}
\newcommand*{\cR}{\con{R}}
\newcommand*{\cD}{\con{D}}

\newtheoremstyle{mytheoremstyle} 
{6pt}               
{6pt}                    
{\itshape}                 
{}					
{\bf}               
{}                       
{.5em}                      
{}  
\theoremstyle{mytheoremstyle}
\newtheorem{satz}{Satz}[section]

\newtheorem{remark}[satz]{Remark}
\newtheorem{algorithm}{Algorithm}[section]
\theoremstyle{remark}
\newtheorem{ass}[satz]{Assumption}

\graphicspath{{./../}}

\numberwithin{equation}{section}
\usepackage{cleveref}

\definecolor{ballblue}{rgb}{0.1, 0.67, 0.7}

\title[Rule-Based Epidemiological Framework]{A Rule-Based Epidemiological Modelling Framework}

\author[D. Alonso]{David Alonso}
\address{Theoretical and Computational Ecology, Center for Advanced Studies of Blanes (CEAB-CSIC),
Spanish Council for Scientific Research, Acces Cala St. Francesc 14, E-17300 Blanes, Spain}
\email{dalonso@ceab.csic.es}

\author[S. Bauer]{Steffen Bauer}
\address{Interdisciplinary Center for Scientific Computing, Ruprecht-Karls-Universit\"at Heidelberg,
Mathematikon,
Im Neuenheimer Feld 205, 69120 Heidelberg, Germany}
\email{steffen.bauer@iwr.uni-heidelberg.de}

\author[M. Kirkilionis]{Markus Kirkilionis}
\address{Warwick Mathematics Institute, University of Warwick, Coventry CV4 7AL, United Kingdom}
\email{corresponding author: mak@maths.warwick.ac.uk}

\author[L. M. Kreusser]{Lisa Maria Kreusser}
\address{Department of Mathematical Sciences
University of Bath, Bath, BA2 7AY, United Kingdom}
\email{lmk54@bath.ac.uk}

\author[L. Sbano]{Luca Sbano}
\address{Liceo “Vittoria Colonna”, Rome, Italy}
\email{luca.sbano@scuola.istrizione.it}

\begin{document}

\begin{abstract}  
Motivated by chemical reaction rules, we introduce a rule-based epidemiological framework for the systematic mathematical modelling of future pandemics. Here we stress that we do not have a specific model in mind, but a whole collection of models which can be transformed into each other, or represent different aspects of a pandemic, and these aspects can change during the course of the emergency, as happened during the Covid-19 pandemic. As conditions for outbreaks in the modern world change on different time-scales, some rapidly, epidemiology has few 'laws', besides perhaps the fundamental infection process described by Kermack-McKendrick \cite{Kermack:1927vk}. Each single of our variety of models, called framework, is based on a mathematical formulation that we call a rule-based system. They have several advantages, for example that they can be both interpreted stochastically and deterministically, without changing the model structure. Rule-based systems should be easier to communicate to non-specialists, when compared to differential equations. Due to their combinatorial nature, the rule-based model framework we propose is ideal for systematic mathematical modelling, systematic links to statistics, data analysis in general and also machine learning leading to artificial intelligence. 
\end{abstract}   

\maketitle

\section{Introduction}
\label{sec:intro}

This paper is a comprehensive study to apply novel mathematical modelling methods after experiencing the COVID-19 pandemic. We propose a rule-based epidemiological framework and hope to use this approach to solve  the still-existing prognosis problem (see Section \ref{sec:discussion:subsec:prognosis}), which is a general problem for most so-called complex systems \cite{Ladyman:2020wf}. To summarise what the reader can expect, here is a list of novelties (some of which we aim to include in the future) of our rule-based framework:

\begin{description}
\item[Formal Language] We use a formal mathematical language here which we call \emph{rule-based} as a foundation of our mathematical models. Rule-based models rely on an extension of reaction kinetics which was first formulated to describe chemical reactions. It has the advantage that the rules can be interpreted both stochastically and deterministically. Moreover the rule-based systems have a mathematical structure which makes them very well suited for both science communication and rigorous mathematical analysis at the same time. We can both derive  stochastic processes and differential equations to describe the temporal dynamics of the pandemic and look for differences. There are several examples in  Section \ref{sec:standardmodels} where certain qualitative properties of the models with stochastic or deterministic interpretation stay invariant under update changes. In these cases the deterministic interpretation should be favoured, as the qualitative for differential equations is very well developed. However, we also found many cases where there is a difference between the stochastic and the deterministic interpretations of the rules differ. Therefore, to avoid ambiguity, we use the stochastic case as the generic simulation approach.
\item[Mathematical Analysis \& Simulation (Scientific Computing)] We use different ways of understanding our rule-based mathematical models, and believe only in this combination we do really understand a mathematical model. A mathematical analysis, most often based on the deterministic interpretation as a differential equation, has the advantage to understand a relatively simple model more or less completely. Much in the same way a climate model can only be understood by a whole range of simulations, more complex models of a pandemic can only be understood by means of exhaustive simulation. The models in this publication are still quite modest, and we will go new ways of scientific publication by making the simulation scripts available to the public.
\item[Model Variation] We do not believe that in something as complex as the COVID-19 pandemic there is a single set of equations or code, like in individual based approaches, that can successfully predict the time course of the pandemic in some sense over a longer period, i.e. overcome the prognosis problem. We stick to the notion that a mathematical model should try to be minimal, i.e. minimally complex by some complexity measure \cite{Ladyman:2020wf} as long as it is sufficiently predictive. This is related to the scientific principle called Occam's razor. As in \cite{MacKay:2003wc}, Chapter 28, we consider Occam's razor as a natural outcome of coherent interference based on Bayesian probability.  In reality this reasoning will imply the optimally predictive model will change, i.e. we will have to walk through a model space, see Section \ref{sec:intro:subsec:framework}.
\item[Data Science] The mathematical modelling community has too long neglected the data aspect. The COVID-19 pandemic has produced a massive amount of data, see Section \ref{sec:discussion:subsec:data}. The model variation and prognosis problem can only be solved together with a firm link to data science. This is a major step we like to tackle in this project in future, however this is not an easy task. We are aware of recent advances in parametrisation of mathematical models, but still believe this is largely an unsolved problem of immense importance.
\end{description}

\subsection{Challenges Triggered By Past Pandemics}
\label{sec:intro:subsec:challenges}

The COVID-19 pandemic posed a complex set of questions to the mathematical modelling community, with nearly daily new twists and turns. This paper intends to present a systematic framework of possible models that are designed to give answers to different aspects triggered by COVID-19 pandemic observations, i.e. empirical epidemiological data. Therefore, as discussed, we are not focusing on a particular model, as is often the case in applied mathematics papers, but follow a systematic model construction approach instead. In this first publication, we do not parametrise our models with data yet, but take basic empirical facts from the data to make \emph{variable classification} choices. To have a first model choice skeleton, we observe the following fundamental empirical facts about COVID-19: 

\begin{description}
\item[Age] From the start of observed coronavirus cases, the associated risks have strongly been age correlated \cite{Palmer2021,ODriscoll2021,Mallapaty2021}.
\item[Health Systems] COVID-19 has put unprecedented pressure on healthcare systems worldwide. Due to the limited capacity of the national health systems or similar institutions, fatalities increased dramatically \cite{ElBcheraoui2020,Haldane2021}. 
\item[Locations] Numerous observations and simulation hint at the risk of infection being location dependent. Staying in closed rooms close to infected individual creates a higher risk of infection when compared to the same arrangement in open space \cite{Wodarz2021}. We leave the investigation of location structure to future publications.
\item[Immunity] Immunity is now in the discussion in connection with the effectiveness of different vaccines, but it has a more general importance. For example, how long are individuals immune to new infections? Can immunity be gained in other ways, other than having had an infection or getting vaccinated? \cite{Gaebler2021,Danchin2021}
\item[Infectivity] Infectivity is usually thought to be rather constant, only depending on outer conditions, such as location, i.e.\ closed narrow rooms are considered dangerous. But clearly infectivity follows a certain time course after an individual gets infected. On top of this, there is indication some individuals, perhaps depending on genetic variation, have a strong infectivity, called 'super-spreaders' \cite{Lewis2021,Wolfel2020}. However, according to our knowledge, there is no clear data source for this yet. 
\item[Behaviour] Given the location dependency of coronavirus infections, it became clear that behavioural choices by individuals play a crucial role in pandemic spread \cite{Jentsch2020}. 
\item[Mobility] Mobility was identified as a major risk factor  from the beginning of the COVID-19 pandemic, as is the case for most individual to individual infection processes \cite{Jentsch2020,Heroy2021,Nouvellet2021}. 
\item[Lockdowns] The age and location dependency of the coronavirus infection rate implied the only successful policy to combat the pandemic was to isolate vulnerable parts of society, and close places where people would get crowded together. A general decrease of mobility would also decrease risk of infections \cite{Lavezzo2020}. 
\item[Testing] Observed time series report tested individuals. Testing is the only way to know how much infection is out there. There is then a fundamental difference between the intrinsic dynamics of total infected individuals, and how many of these individuals we know they are really infected (or not) because they are counted and reported through testing. From a modelling perspective, this requires to include new dynamic variables or sub-types to follow the time evolution of how many individuals tested positive (or negative), and for how long a test is still reliable \cite{Mercer2021}. 
\item[Vaccination] The rapid development of vaccines gave for the first time rise to the hope that a sufficiently large percentage of populations could be immunised safely to stop the spread of the disease (herd immunity) \cite{Mallapaty2021a,Wang2021}. 
\item[Mutations] These are extremely important at the moment of disease emergence \cite{Dobson2015}, but also the latest twist of the COVID-19 pandemic was the rapid emergence of new mutations of the virus, with some variants observed to be more infectious, and deadlier \cite{Wang2021}. 
\end{description}

All of these items in the discussion above can only be incorporated in mathematical models after a scientific classification. Using these models the future of  pandemics can be investigated, including the likelihood of major new outbreaks. The likeliness of viruses like the coronavirus crossing from animals like bats to humans is largely related to habitat and whole ecosystem destruction \cite{Hassell:2021uu}. Is it possible to achieve herd immunity? How will herd immunity  be influenced by future mutations with different infectivities? Will COVID-19 be staying forever somewhere in the world, or will we be able to eradicate it with our policy measures, and vaccination programmes \cite{Aschwanden2021}? 

In this paper we only consider \emph{discrete} classifications, see Section~\ref{sec:intro:subsec:classification}. After the classification the resulting types will become variables in the mathematical model. However, becoming a variable is not the only possibility to project a dependency of above concepts into a mathematical model. We will introduce rates, a fundamental concept of any dynamic, time-dependent model, and we will also use types which results in a functional dependency of rates at which events occur, such as infections, on the system state.

\subsection{Classification as Structural Model Basis, Relation to Statistics}
\label{sec:intro:subsec:classification}

To understand our framework model approach, we must first discuss classifications, where objects are grouped into types. Following this idea, we arrive at a simple modelling question: what are the best types or classes\footnote{later becoming model variables, see Section \ref{sec:intro:subsec:dynamics:sub:mc}.} which when used to group human individuals serving as basic objects of a mathematical model, can successfully establish a predictive model of the COVID-19 pandemic on the basis of these group interactions? The grouping of individuals we call a type, because all individuals of the group must have a common property, i.e.\ a type. 

We would like to emphasize that it is the type that is the basic entity that connects our mathematical model to empirical data. In this context the type can be interpreted as a random variable, and the grouping of individuals according to types corresponds to random experiments. Sometimes the random experiment is easy to perform. For example, if we consider a certain population and choose age classes as types, we can simply hand out questionnaires. However, if our type is something like `immunity', elaborate lab work is needed which very likely can only be measured in random samples, i.e. sub-populations. 

\begin{figure}[htbp] 
   \centering
   \includegraphics[width=8cm]{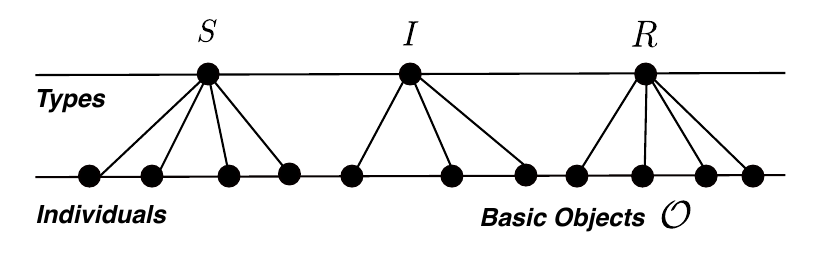} 
\caption{The object set for humans based epidemiology are usually individuals. Here, we have three types in the classical classification of human individuals, susceptible ($S$), infected ($I$) and recovered ($R$) individuals. This will be the basic classification of all models discussed, with additional classifications on top of this structure. An edge from individuals to the types, here disease states, indicates the individual are of this type.}
   \label{fig:sec:model:subsec:types}
\end{figure}

The most important type of an individual in this context is the characterisation that an individual is `infected', establishing the type $I$ as an example. More general, any epidemiological model considers interacting populations of finite-structured types of individuals which we can count inside some system boundaries, like a national or regional territory. Let $\mathcal{O}$ be the set of \emph{basic objects}, here human individuals of a larger population. The individuals form the system components at the micro-level. In principle, we would need to let $\mathcal{O}$ to be parametrised by time, because we could allow the set to change when objects either enter or leave the system. But in this description of the pandemic, we assume the population size is fixed, therefore no individuals are allowed to 
`enter' or `leave'. This means, if we introduce a type $D$ for `dead' later on, we still count individuals of this type to be part of the population. The single layer classification of an $SIR$-model is shown in Figure \ref{fig:sec:model:subsec:types}. 

The types are also interpreted as a scientific classification of individuals. For each classification we require that an individual is exactly of one type. However, there are possibly many different relevant classifications of individuals, such as age, or disease state etc., and these can exist in parallel. In addition, individuals can be of a certain immunity type, and they can have a certain infectivity, also a type. Let $\mathcal{T}=\{T_1,\ldots, T_s\}$, $s \in \mathbb{N}$, denote the set of all possible types or individual characteristics, $s$  different types in total. It follows that we have created a hypergraph $H = H(\mathcal{O},\mathcal{T})$, i.e.\ the types form a set of hyperlinks, which are non-overlapping. Classifications like age or location can be interpreted as either internal or external states of an individual, and have very likely so-called sub-types. The best example is age, where the number of chosen age classes form the sub-types of the age type, i.e.\ the classification by age. The state of the system is given by the number of basic objects of some type, here number of human individuals that have a type at some time $t$. We use the notation that the number of these individuals of type $T_i$ is given by $n_i(t) \in \mathbb{N}_0$ for all $i=1,\ldots,s$. 

Of course, it is possible that there is more than one system \emph{compartment} in which objects can reside. In this case the types are indexed by compartment, making them compartmental sub-types. An obvious generalisation of the situation discussed here are \emph{multiple compartments}, and such models will have to incorporate movement between compartments. Multi-compartment models can be used to consider \emph{geographical} models of the epidemic. We leave this for future publications.

\subsection{Rules}
\label{sec:intro:subsec:classification:sub:rules}

As a next step we introduce rules (or reaction channels). Let $\mathcal{R}=\{R_1,\ldots, R_r\}$, $r \in \mathbb{N}$,  be the finite set of rules constituting the epidemiological system.  Each rule $R_j$ with $j=1,\ldots,r$, takes the form
\begin{eqnarray}
\label{eq:sec:model:subsec:reaction:reactionscheme}
    \sum_{i=1}^{s} \alpha_{ij}  T_i  \xrightarrow{k_j} \sum_{i=1}^{s} \beta_{ij}  T_i,
\end{eqnarray}
where $k_j$ denotes the rate constant of rule (reaction) $R_j$. The sums have to be understood as so-called formal sums (as usual in reaction systems). The coefficients $\alpha_{ij} \in \mathbb{N}_0$ are the stoichiometric coefficients of types $T_i$, $1 \le i \le s$, of the \emph{source} side, and  $\beta_{ij}\in \mathbb{N}_0$ are the stoichiometric coefficients of types $T_i$ of the \emph{target} side of  transition or reaction $R_j$. These stoichiometric coefficients describe changes of individual numbers due to events, such as an infection. Note that either all stoichiometric coefficients on the source side, or all stoichiometric coefficients on the target side are allowed to be zero. In this case, the formal sum is replaced by the empty set  $\emptyset$. 

Note that also rules are subject to confirmation by empirical data. In this case the random variable is the rule itself, and the random experiment consists of observing such a transition in the population. For example, it has never been observed that more than one infected person is needed to infect a susceptible person in the COVID-19 pandemic, therefore the stoichiometric coefficients are also observables, i.e. random variables linked to experiments.

\subsubsection{Global Information}
\label{sec:intro:subsec:classification:sub:global}

Rules are generalisations of reactions as used in so-called reaction kinetics. In epidemiology, but also other areas of science, there is a need to additionally introduce \emph{global information} into the system description. The idea is that individuals know, for example from the daily news, how many infected people there are in their area, and possibly change their behaviour accordingly. Note that in chemical reaction systems, such global information is completely missing, the atoms or molecules follow only local information, i.e.\ they experience bumping into each other, but they do not know anything about the reaction partners before a reaction event. 

We assume that the global information is collected from the type set $\mathcal{T}=\{T_1,\ldots, T_s\}$, $s\in\N$. In this context  $\mathcal{T}$ will be referred to as \emph{local types}. We introduce the \emph{global information operator} $\Delta$ and a \emph{global information type set}, or short \emph{global types}, given by $\mathcal{G}=\{G_1,\ldots, G_g\}$, $g \in \mathbb{N}$. Now let
$$ \Delta \colon \mathcal{T} \to \mathcal{G}. $$
Therefore, $\Delta$ takes information from the types of the model, and maps it to global information which can be made known to the system. For example, in an age-structured model, let $\mathcal T_I=\{ I_1, I_2, \ldots, I_n\}$ denote the types of infected individuals inside $n$-different age classes. Let $\mathcal{G}=\{G_I\}$ be a one-dimensional global information type set, here interpreted as all infected individuals in the system. Then $\Delta$ is given implicitly by the formal sum
$$ G_I := \sum_{i=1}^{n} I_i. $$
In general we assume
$$ G_i := F_i(\mathcal{T}), \;\;\; i=1,\ldots,g, $$
with $F_i$ some function mapping $s$ many types to one type. When we introduce a state  $n_i(t) \in \mathbb{N}_0$ of type $T_i$ at system time $t\geq 0$ for $i=1,\ldots,s$, see Section \ref{sec:intro:subsec:dynamics:sub:mc}, then global types will at the same time have derived numerical values.
%
\subsubsection{Sub-Types as Sub-Classifications}
\label{sec:intro:subsec:classification:sub:types}
\begin{figure}[ht]
    \centering
    \includegraphics[width=0.8\linewidth]{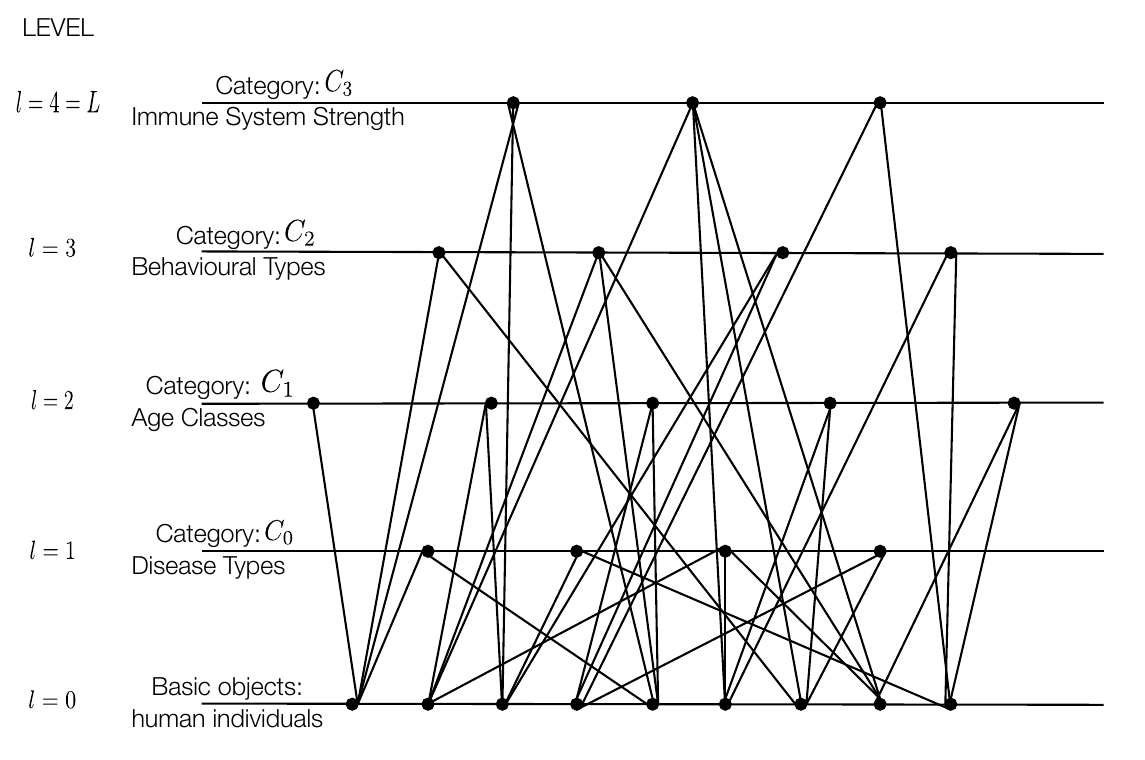}
    \caption{Different categories structuring a human population into epidemiological modelling relevant categories. Shown is a simple population sketch with individual number $N=9$ and frozen time. At each observation time each individual is exactly member of one type inside each category, i.e. we assume complete partition. This implies the summation of degrees of types in each category must be the population number $N$. The degree of each individual must be the number $L$ of categories introduced.}
    \label{fig:epi_mod_cat}
\end{figure}

%
In many epidemiological applications so-called sub-types play a crucial role. They are related to multi-category models, with the simplification that a certain rule-set is regarded as the base model, and then further structured with one or more additional classifications, see Figure \ref{fig:epi_mod_cat}. Certain classifications might only be needed at certain times, for example, mobility patterns might change during a pandemic. This is a strong argument never to rely on a particular model throughout a pandemic, epidemiology is better understood to be a complex system. In this paper the SIR-model is used as the basic model, and then all further modelling is done via the sub-type modelling approach. The best example in our current context are age classes, see model (\ref{eq:sec:standard:subsec:age-sird}) in sub-type notation. 
%
%
Because types can now exist of different sub-types, we can introduce their frequency distribution. Define $l_i \in \mathbb{N}$ as the number of sub-types of type $i$ for $1 \le i \le s$. Let $\mu_{ij}^\alpha$ denote the frequency of the sub-type $\alpha$, $1\le \alpha \le l_i$, needed for the source terms in rule  $R_j$, called the {\bf source frequency}. We have $$\mu_{ij}^\alpha =  \frac{n_{ij}^{s,\alpha}}{m^s_{ij}},$$ where $n_{ij}^{s,\alpha} \in \mathbb{N}$ is the number of objects (individuals) of type $i$ and sub-type $\alpha$ needed for the source terms in rule  $R_j$, and $m^s_{ij} := \sum_{\alpha = 1}^{l_i} n_{ij}^{s,\alpha}$ is the total number of objects of type $T_i$ needed for reaction $R_j$.  The symbol $s$  denotes `source'. Clearly, we  have 
\begin{equation}
\label{eq:sec:model:subsec:reaction:frac:1}
\sum_{\alpha=1}^{l_i} \mu_{ij}^\alpha = 1 \; \; 
\end{equation}
for all types  $i=1,\ldots,s$ and all reactions $R_j$ for $j=1,\ldots,r$. 
Similarly, let $\nu_{ij}^\alpha$ denote the frequency of the sub-type $\alpha$, $1\le \alpha \le l_i$ after rule  $j$ has been triggered, called the {\bf target frequency}. We have $$\nu_{ij}^\alpha =  \frac{n_{ij}^{t,\alpha}}{m^t_{ij}},$$ where $n_{ij}^{t,\alpha} \in \mathbb{N}$ is the number of objects (individuals) of type $i$ of sub-type $\alpha$ as result of the event described by rule  $R_j$, and $m^t_{ij} := \sum_{\alpha = 1}^{l_i} n_{ij}^{t,\alpha}$.  Here the symbol $t$ denotes `target'. Again we have
\begin{equation}
\label{eq:sec:model:subsec:reaction:frac:2}
\sum_{\alpha=1}^{l_i} \nu_{ij}^\alpha = 1 
\end{equation}
for all types  $i=1,\ldots,s$ and all reactions $R_j$ for $j=1,\ldots,r$.
The constants $\mu_{ij}^\alpha,\nu_{ij}^\alpha\in \mathbb{Q}$ specify the frequency of individuals that are necessary for each rule to be realised. They are therefore the frequency analogues of stoichiometric coefficients. Finally we can now write every possible event or rule (or reaction) $R_j$ in the more general form
\begin{eqnarray}
\label{eq:reactionscheme:generalised}
\sum_{i=1}^{s} \gamma_{ij} \sum_{\alpha=1}^{l_i}  \mu_{ij}^\alpha T^\alpha_i  \xrightarrow{k_j} \sum_{i=1}^{s} \delta_{ij} \sum_{\alpha=1}^{l_i}  \nu_{ij}^\alpha T^\alpha_i,
\end{eqnarray}
for $j=1,...,r$, where the sums have to be understood again as  formal sums. The coefficients $\gamma_{ij} \in \mathbb{N}$ are the stoichiometric coefficients of types $T_i$, $1 \le i \le s$, of the source side, and  $\delta_{ij}\in \mathbb{N}$ are the stoichiometric coefficients of types $T_i$ of the target side of rule $R_j$ with rate constant $k_j$. These stoichiometric coefficients describe changes of object numbers, independent of the sub-type of an object of any type. Note that if all types have just a single state or configuration, then the scheme is equivalent to a traditional reaction scheme (\ref{eq:sec:model:subsec:reaction:reactionscheme}) by construction. The generalised scheme (\ref{eq:reactionscheme:generalised}) is useful, for example, we can use it to establish a varying number of discrete age classes in a single rule. For multi-category models, not considered here. a more complex notazion based on tensors, is needed. Of course sub-types can also be used to introduce multi-compartment, i.e. geographical meta-population models, but a too naive interpretation can lead to modelling inconsistencies.
%
%
%
\subsection{Towards an Epidemiological Modelling Framework}
\label{sec:intro:subsec:framework}
%
We are now ready to establish our modelling framework, which  by construction will become a discrete and finite model space. As we will argue, this is important as we later like to apply machine learning technology. A discrete model space will be more easily searchable by automated methods, comparing their predictions with data. However, there needs to be a fundamental discussion on whether we can construct such a discrete and finite model space which must incorporate models that are \emph{predictive}, otherwise the construction would be useless. Moreover we must fix an interpretation of the identity of a model, i.e.\ answer the question when two models are different? All these questions will occupy us in future publications. We summarize our findings in the following list:
\begin{description}
\item[Structural (Mathematical) Ontology] When are two models different? We believe this depends on the kind of question being asked. The two possibilities are: (i) the model structure only decides whether two models are different, independent of any empirical association, or (ii) two models become different if they are allowed to have identical structure, but different empirical interpretation. Following (i), we can mathematically investigate a model, for example the qualitative behaviour, which will be the same independent of any empirical association. This is the usual mathematical approach. For example, considering simplicial complexes in algebraic geometry, the same simplex can occur several times in a complex, see \cite{Jost:2015vk}, page 70. Indeed, if we construct models, for example on basis of rule-based systems, we might like to define algebraic operations on models, adding or subtracting types and rules in order to derive new models. This will become similar to the concept of a $q$-chain, the starting point of the homology of simplicial complexes, just with replacing $q$-simplices with models having a certain number of types and rules. See \cite{Jost:2015vk}, page 179. Such a concept we refer to as \emph{structural ontology}.
\item[Empirical Ontology] When are two models different, based on empirical data? It is best to understand this question by an example. Consider the model

\begin{equation}
\label{eq:sec:intro:subsec:framework}
\begin{alignedat}{4}
	T_1 + T_2 &\xrightarrow{\alpha} && 2 T_2 \\
	      T_2 &\xrightarrow{\beta}  && T_3\\
\end{alignedat}
\end{equation}

with $3$ types and $2$ rules. If we set $(T_1,T_2,T_3) = (S,I,R)$, then we recover the simplest infection model (\ref{eq:sec:standard:subsec:sir}). Setting the types in this way makes a connection to empirical data, as from now onwards we can make measurements in a given population, for example to check whether this model is predictive. By setting $T_1$ to a certain well-defined opinion, $T_2$ to another well-defined opinion, and $T_3$ to, say, no opinion about the subject at all, i.e. being indifferent, this makes (\ref{eq:sec:intro:subsec:framework}) an opinion formation model. As this model would be evaluated according to very different data, it would make the two models empirically different. We call this interpretation \emph{empirical ontology}.
\end{description}

\section{Modelling Framework}
\label{sec:framework}

We are now able to present our epidemiological framework in more detail. We first look briefly at the connection to structural (static) modelling.

\subsection{Introducing Dynamics}
\label{sec:intro:subsec:dynamics}

\subsubsection{Continuous-Time Markov Chains Including Global Information}
\label{sec:intro:subsec:dynamics:sub:mc}

In this section, we formulate reactions \eqref{eq:sec:model:subsec:reaction:reactionscheme} as a continuous-time Markov chain. We regard the size/number $n_i(t)  \in \mathbb{N}$ of type $T_i$ at system time $t\geq 0$ for $i=1,\ldots,s$ as discrete random variables with values in $\{0,\ldots,N\}$.  For defining the transition probabilities associated with the stochastic process we assume that the process is time-homogeneous (autonomous), i.e.,  while no event occurs, the transition probabilities remain constant, and depend only on the time between events $\Delta t$, but not on the specific time $t$. We observe the system from time $0$ up to time $T>0$. In addition, we suppose that the Markov property is satisfied, that is, the probability distribution of a future state of the stochastic process at time $t+\Delta t$ only depends on the current state at time $t$, but not on any state prior to $t$.  With each type $T_i$ having an associated state $n_i(t)  \in \mathbb{N}$, the global type set $\mathcal{G}=\{G_1,\ldots, G_g\}$, see Section \ref{sec:intro:subsec:classification:sub:global}, get derived numerical values which are. We assume at system time $t\geq 0$ to obtain the \emph{global state vector} $\theta(t) := (\theta_1(t), \ldots, \theta_g(t))$, with $\theta_i \in  \mathbb{R}$ for $1 \le i \le g$. Given the system is in state (or configuration) $n$, we introduce the propensities $\kappa_j(t^\prime,n)$ for $j=1,\ldots,r$. Here, the propensity $\kappa_j(t^\prime, n)$ of the $j$th reaction $R_j$ is the probability that the $j$th event/reaction occurs within an infinitesimal time interval, the last event in the entire system happened at $t^\prime$, and is given by 
\begin{align}\label{eq:propensity}
 \kappa_j(t^\prime, n) = k_j f_j (\theta(t^\prime,n))  \prod_{i=1}^s   \begin{pmatrix} n_i \\ \alpha_{ij} \end{pmatrix},
\end{align}
where $\kappa_j(\theta(t^\prime))$ denotes the global state dependent propensities of reaction $R_j$, $k_j$ is the reaction constant, and the combinatorial factor reflects the number of ways in which reaction $R_j$ may happen, see (\ref{eq:sec:model:subsec:reaction:reactionscheme}). The functions $f_j$, $1 \le j \le r$, are called \emph{global information functional responses}. These functions modify the rule execution constants $k_j$ according to the state of the given global information vector $\theta$.

\begin{remark}
Note that the notation $\kappa_j(t^\prime, n)$ suggests time-dependence, which is not the case, therefore simply writing $\kappa_j(n)$ is equally valid and will be adopted sometimes. The notation is just a reminder that the propensity depends on the last event that happened in the system, as then the global information available to the system is updated. Time-homogeneity and the Markov property are still valid piecewise. In more sophisticated situations however, such as externally imposed lockdowns, the time-dependency of the propensity $\kappa_j$ will have to be considered explicitly.
\end{remark}

We use some standard algebraic expressions for global information functional responses $f_j(\theta)$ in case $\theta$ is one-dimensional and non-negative:

\begin{description}
\item[Simple Saturation] $$ f_j(\theta) := \frac{1}{1+ \lambda_j \theta}. $$ Here $\lambda_j$ is an additional positive parameter. We also consider $ \frac{1}{f_j}$ in case the functional response has to be monotonically decreasing, due to modelling purposes.
\item[Michaelis-Menten] $$ f_j(\theta) := \frac{1}{\lambda_j + \theta}. $$
Here $\lambda_j$ is an additional positive parameter. Alternatively, we may consider $ \frac{1}{f_j}$ in case the functional response is monotonically decreasing.
\item[Exponential] $$ f_j(\theta) :=  \lambda_j e^{-\lambda_j \theta}. $$
Here $\lambda_j$ is an additional positive parameter. The exponential functional response gives rise to a probability measure $P_j$, with distribution function $F_j (\theta) = 1 -  \lambda_j e^{-\lambda_j \theta}$, and is called \emph{exponential distribution}. For $\epsilon>0$, we define a loss rate $\lambda(\theta)$ by
$$ \lambda(\theta) := \lim_{\epsilon \to 0} \frac{P((\theta,\theta + \epsilon])}{P((\theta,\infty))}. $$
It follows that
$$ \lambda(\theta) =  \frac{f_j(\theta)}{1-F_j (\theta))}. $$
\item[Weibull] $$ f_j(\theta) :=   \lambda_{j,1} \lambda_{j,2} \theta^{ \lambda_{j,2} - 1} e^{- \lambda_{j,1} \theta^{\lambda_{j,2}}}. $$
Here $\lambda_{j,1}, \lambda_{j,2}$ are  positive parameters. Note that for $\lambda_{j,2}=1$, $f_j$ reduces to the exponential distribution. 
\end{description}

If the global state vector $\theta(t) := (\theta_1(t), \ldots, \theta_g(t))$ has dimension $g>1$, then each of the global information functional responses is assumed to be given in a multiplicative way by $g$-many sub-functionals, each dependent on one component of $\theta$:
$$ f_j (\theta) = f_{j,1}(\theta_1)f_{j,2}(\theta_2)\cdots f_{j,g}(\theta_g), $$
where each  $f_{j,i}$ is given by one of the one-dimensional algebraic expressions given above. From a modelling point of view, this assumes different global information states have an independent influence on the event rate of a rule, such as an infection. An example would be avoidance and mobility. Avoidance can be modelled by a monotonically decreasing functional response, mobility as a monotonically increasing functional response. Now, based on the assumption that at most one reaction can occur within an infinitesimal time interval $\Delta t$, the transition rates for the Markov process, given that the system is in state $n$, can then be approximated as $\kappa_j(n)\Delta t$ for reaction $R_j$ to occur within time interval $\Delta t$ and $1-\sum_{j=1}^r \kappa_j(n)\Delta t$ for no reaction to occur within time interval $\Delta t$. To simulate the sample paths of the continuous-time Markov chain, one can apply the Gillespie algorithm. The simulation of many trajectories of the system  allows the computation of the statistics of evolution.

\subsubsection{Gillespie's Algorithm}
\label{sec:intro:subsec:dynamics:sub:gillespie}

The most important stochastic update idea for epidemiological models is \emph{Gillespie's Stochastic Simulation Algorithm (SSA)} and its generalisation. It is the main algorithm used to study the stochastic evolution of the models presented in this paper. This evolution arises due to the interactions among types $\mathcal{T}$ and, in the language of reaction schemes, can be expressed  as a reaction network. Reaction $R_j$ is given by Equation (\ref{eq:sec:model:subsec:reaction:reactionscheme}) for $j=1, \ldots, r$, where $k_j$ denotes the rate constant of reaction $R_j$. Interpreting a reaction as a rule, we will use the terms \emph{reaction} and \emph{rule} interchangeably. Each process occurs with a \emph{propensity} $\kappa_j(n)$, defined as
\begin{equation}
 \kappa_j(n) := k_j \, h_{j} (n),
\label{eq:Rj}
\end{equation}
where  
$$ h_j(n) := \prod_{i=1}^s   \begin{pmatrix} n_i \\ \alpha_{ij} \end{pmatrix}.$$
Here, the combinatorical factor $h_j$ describes the number of ways in which reaction $R_j$ takes place. As in the original paper \cite{Gillespie1976} by Gillespie, we make the following assumption:

\begin{ass}
\label{Gillespie_assumption}
Reaction $R_j$ occurs within a time interval $\Delta t$ with probability $\kappa_j(n)\Delta t$.
\end{ass}

For a given population with system configuration $n \in \mathbb{R}^s$, one can  show (see \cite{Gillespie1976}) that the probability density associated with the occurrence of $R_j$ is given by
\begin{equation}
\Phi_j(n,t)=\kappa_j(\theta(n),n)\exp\left(-\sum_{l=1}^r\kappa_l(\theta(n), n)\,t\right)
\label{eq:gillespie_prob_Pi}
\end{equation}
as a function of time $t$. The probability that reaction $R_j$ ever occurs is given by 
\begin{equation}
\rho_j(n)=\int_0^\infty \di t'\,\Phi_j(n,t')=\frac{\kappa_j(\theta(n),n)}{\sum_{l=1}^r \kappa_l(\theta(n),n)}
\label{eq:rho_i_standardgil}
\end{equation}
for  $j\in\{1,...,r\}$. The rules given by Equation (\ref{eq:sec:model:subsec:reaction:reactionscheme})  generate a stochastic process which satisfies the Markov property due to Assumption \ref{Gillespie_assumption}. This process is  the starting point for the derivation of the master equation, describing the time evolution of the probability distribution of the system as a whole (see \cite{Gardiner}). The probability density in \eqref{eq:gillespie_prob_Pi} can  be interpreted  in terms of reaction waiting times and, in fact, instead of Assumption \ref{Gillespie_assumption}, we can equivalently assume that  reaction $R_j$ occurs with an \emph{exponentially distributed} reaction waiting time. In \cite{Gillespie1976} Gillespie showed that the master equation can be exactly solved through the construction of realisations 
of the stochastic process associated to \eqref{eq:sec:model:subsec:reaction:reactionscheme}.    This is achieved through the following algorithm, where $::=$ denotes replacement: 
\begin{algorithm} 
\label{alg:sec:model:subsec:gillespie}
{\bf [Gillespie Algorithm]} 
\begin{description}
\item[0] Initialize the starting time $t$, the system configuration $n=(n_1,\ldots,n_s)$ and the rate constants $k_j$ for $j=1,\ldots,r$. 
\item[1] Generate two random numbers $r_1, r_2$ uniformly distributed in $[0,1]$.
\item[2] Set $\Phi:= \sum_{l=1}^r \kappa_l (n)$ and compute $\tau:= -\frac{1}{\Phi} \ln r_1$.
\item[3] Set $t::=t+\tau$ as the time of the next rule execution.
\item[4] Determine the reaction $R_i$ which is executed at time $t$ by finding $i$ such that
\begin{align}\label{eq:conditiongillespie}
 \frac{1}{\Phi} \sum_{l=1}^{i-1}\kappa_l (n)<   r_2 \leq \frac{1}{\Phi} \sum_{l=1}^{i}\kappa_l (n).
\end{align}
\item[5] Execute rule $R_i$ and update the new system configuration $n$.
\item[6] Go to step {\bf 1} if $t<T$, otherwise stop.
\end{description}
\end{algorithm}

The random numbers generated in Algorithm \ref{alg:sec:model:subsec:gillespie} are used for determining the index $i$ of the next rule to be triggered, and the time interval $\tau$ for reaction $R_i$ to occur. By the definition of $\rho_l(n)$ in \eqref{eq:rho_i_standardgil} we have
$$\frac{1}{\Phi} \sum_{l=1}^i \kappa_l(n)=\sum_{l=1}^i \int_0^\infty \di t'\,\Phi_l(n,t')=\sum_{l=1}^{i}\rho_l (n)$$
in step {\bf 4} in Algorithm \ref{alg:sec:model:subsec:gillespie}. We use Gillespie's algorithm to simulate the model dynamics, based on the assumption that the processes have exponentially distributed waiting times or, equivalently, that they are all Poisson processes, like atomic or molecular mixtures of gases or liquids. In complex systems formed by types which are heterogeneous and evolve through diverse interactions that cannot be always characterised by \emph{exponentially distributed waiting times}, therefore the Gillespie approach is not fully adequate. It is thus very important to have a  generalisation of Gillespie's algorithm that allows us to study systems driven by processes with different waiting time distributions and not necessarily exponentially distributed waiting times. Several generalisations exist in the literature, see  \cite{CCTRW}, \cite{Gillespie_ME_with_waiting_times}, \cite{Masuda} and \cite{Bog} for instance, where a generalised Gillespie algorithm  based on a generalised master equation in the form of an integro-differential equation \cite{CCTRW} may be considered. 

\subsubsection{Kolmogorov Differential Equations and Master Equations}
\label{sec:intro:subsec:dynamics:sub:kolmogorov}

Based on the propensities $\kappa_j(n)$ in \eqref{eq:propensity}, we state the forward Kolmogorov differential equations which are also referred to as master equations and can be used to predict the future dynamics. The Kolmogorov differential equations can be regarded as an alternative way to define the continuous-time Markov chain. We denote by $p_n(t)$ the probability that the system is in state $n$ at time $t$. For determining the time derivative $\frac{\di p_n(t)}{\di t}$, all transitions resulting in state $n$ and all transitions away from state $n$ have to be considered. Under the assumption that at most one reaction can occur within an infinitesimal time interval, we consider  $m^{(j)}=(m^{(j)}_1,\ldots,m^{(j)}_s)$ for $j=1,\ldots,r$ as the transition between states for reaction $R_j$  where $m^{(j)}_i$ denotes the difference of individuals of type $T_i$ before and after reaction $R_j$ for $i=1,\ldots,s$. If the system is in state $n-m^{(j)}$ before reaction $R_j$ occurs, this leads to state $n$ after reaction $R_j$ occurs. We formulate the Kolmogorov equations as
\begin{align}\label{eq:kolmogorov}
    \frac{\di p_n(t)}{\di t}=\sum_{j=1}^r \left( \kappa_j(n-m^{(j)}) p_{n-m^{(j)}}(t) - \kappa_j(n) p_n(t) \right).
\end{align}
Note that \eqref{eq:kolmogorov} can be written as a linear ordinary differential equations (ODEs) of the form 
\begin{align*}
    \frac{\di P(t)}{\di t}=Q P(t)
\end{align*}
for a vector $P$ depending on time $t$ and an appropriate choice of matrix $Q$, containing the probabilities $p_n$ of all states $n$. However, since the number of states of a system is generally very large, this implies that \eqref{eq:kolmogorov} is a very large system of ODEs and hence its numerical solution is computationally expensive. To combine the Kolmogorov equations with some given data (e.g. global information), we introduce a given data input $\mathcal I$ which may depend on time, i.e., $\mathcal{I}=\mathcal I(t)$. We consider the generalised propensity 
$$\mathcal \kappa_j(n,\mathcal I)=g(\mathcal I) k_j  \prod_{i=1}^s   \begin{pmatrix} n_i \\ \alpha_{ij} \end{pmatrix}$$ 
for some nonnegative function $g$. The multiplicative factor $g(\mathcal I)$ connects the probability that the $j$th reaction occurs within an infinitesimal time interval with the known data $\mathcal I$.   The generalised Kolmogorov equations are given by
\begin{align*}
    \frac{\di p_n(t)}{\di t}=\sum_{j=1}^r \left( \kappa_j(n-m^{(j)},\mathcal I) p_{n-m^{(j)}}(t) - \kappa_j(n,\mathcal I) p_n(t) \right).
\end{align*}

\subsubsection{Stochastic Differential Equations}
\label{sec:intro:subsec:dynamics:sub:sde}

The master equation  can be studied by different methods that allow us to compute the deterministic limit of the master equation and its stochastic approximations such as Van Kampen expansions and Moment expansions  which can be justified through the Central Limit Theorem. A general and complete description of these methods can be found in \cite{Gardiner} and \cite{EthierKurtz}.\\
Here we follow a simplified approach to show that  
another stochastic interpretation of the rules from Equation (\ref{eq:sec:model:subsec:reaction:reactionscheme}) leads to \emph{stochastic differential equations}. We have to introduce concentrations $c_i=\frac{n_i}{N}$ as discrete random variables and approximate the master equation (\ref{eq:kolmogorov})  by a Ito's stochastic differential equation. This approximation is the consequence of two facts:

\begin{itemize}
    \item The Kolmogorov equation (\ref{eq:kolmogorov}) is equivalent to a recursion equation for the Poisson processes $\mathcal{P}_j(\zeta)$ for $j=1,\ldots,r$ 
    where $\zeta=\zeta(k,\alpha,c)$ is  a random variable due to the dependence of $\zeta$ on the random variable $c=(c_1,\ldots,c_s)$\footnote{This is the crucial property on which Gillespie's algorithm is based.}.
    \item The Central Limit Theorem for Poisson processes 
     $\mathcal{P}_j(\zeta)$, $j=1,\ldots,r$, allows us to write $\mathcal P_j$ as
    \[\mathcal{P}_j(N\,\zeta)\simeq N\,\zeta + \mathcal{N}_j(0,1)\,\sqrt{N\,\zeta},\]
    where $\mathcal{N}_j(0,1)$ are standard normally distributed random variables and $N$ is the parameter by which $n_n/N\rightarrow c_i$ for $N\rightarrow +\infty$.
\end{itemize}

To describe the stochastic differential equation, 
we introduce the  vector $W(t)=(W_1(t),\ldots,W_r(t))$ of $r$ independent Wiener processes, i.e., $W_i(t)$ is normally distributed with mean zero and variance $t$. We consider the concentration $c_i(t)$ of type $T_i$ at time $t\geq 0$ for $i=1,\ldots,s$ as discrete random variables and write the system of stochastic differential equations as 
\begin{align}\label{eq:sde}
        \di c_i(t)=\sum_{j=1}^r (\beta_{ij}-\alpha_{ij}) k_j  \prod_{l=1}^s c_l(t)^{\alpha_{lj}} \di t+ \sum_{j=1}^r \frac{\beta_{ij}-\alpha_{ij}}{\sqrt{N}} \sqrt{ k_j  \prod_{l=1}^s c_l(t)^{\alpha_{lj}}} \di W_j(t),\quad i=1,\ldots,s,
\end{align}
or, equivalently,
\begin{align*}
    \di c_i(t)=\sum_{j=1}^r (\beta_{ij}-\alpha_{ij}) k_j  \prod_{l=1}^s c_l(t)^{\alpha_{lj}} \di t+ \sum_{j=1}^r \frac{\beta_{ij}-\alpha_{ij}}{\sqrt{N}}\mathcal{N}_j(0,1) \sqrt{ k_j \di t \prod_{l=1}^s c_l(t)^{\alpha_{lj}}},\quad i=1,\ldots,s,
\end{align*}
where $\mathcal{N}=(\mathcal{N}_1,\ldots,\mathcal{N}_r)$ is a vector of $r$ independent standard normally distributed random variables. 
For large $N$ one can neglect the terms proportional to the Wiener processes in \eqref{eq:sde} and this results in \eqref{eq:odesystem}.
The system of stochastic differential equations \eqref{eq:sde} can be solved with the Euler-Maruyama method, a finite difference approximation for stochastic differential equations, given by
\begin{align}\label{eq:sde_disc}
     c_i(t+\Delta t)=c_i(t)+\sum_{j=1}^r (\beta_{ij}-\alpha_{ij}) k_j  \prod_{l=1}^s c_l(t)^{\alpha_{lj}} \Delta t+ \sum_{j=1}^r \frac{\beta_{ij}-\alpha_{ij}}{\sqrt{N}} \mathcal{N}_j(0,1)\sqrt{k_j \Delta t   \prod_{l=1}^s c_l(t)^{\alpha_{lj}}},
\end{align}
for $i=1,\ldots,s$. Note that \eqref{eq:sde_disc} 
is the discrete form of a  stochastic differential equation which is also known as the Chemical Langevin Equation. In the limit $N\to \infty$, we recover the discrete form of the deterministic system \eqref{eq:odesystem}, given by
\begin{align*}
     c_i(t+\Delta t)=c_i(t)+\sum_{j=1}^r (\beta_{ij}-\alpha_{ij}) k_j  \prod_{l=1}^s c_l(t)^{\alpha_{lj}} \Delta t.
\end{align*}

\subsubsection{Deterministic Updates}
\label{sec:intro:subsec:dynamics:sub:deterministic}

Although we are mainly focusing on stochastic dynamics, it is important to compare results with the associated deterministic dynamics, for two reasons. First, the deterministic dynamics analysis reveals a lot of the possible features of an epidemiological model as long as individual numbers are sufficiently large inside each type of the classification, something we could call a practical continuum limit. Secondly, most of the current publications in epidemiology are based on deterministic dynamics, therefore we need to include this possibility for reasons of comparison. 

The deterministic interpretation is given by a system of ODEs which describes the rate of change of size $n_i$ of the types $T_i$ for $i=1,\ldots,s$, over time. Considering the system of reactions \eqref{eq:sec:model:subsec:reaction:reactionscheme},  we can construct the associated stoichiometric matrix of size $s \times r$  and the vector of rate laws. The entry $(i,j)$ of the stoichiometric matrix is given by $\beta_{ij}-\alpha_{ij}$.
In the deterministic setting, we consider the concentration of the $i$th type, given by $c_i=\frac{n_i}{N}$, where the system size $N$ is assumed to be constant. 
The rate law of reaction $R_j$ satisfies $k_j \prod_{l=1}^s c_l^{\alpha_{lj}}$ for $j=1,\ldots,r$, i.e., it is proportional to the powers of the concentrations before the reaction and the deterministic rate constant $k_j$. This leads to the  system of ODEs
\begin{align}\label{eq:odesystem}
    \frac{\di c_i(t)}{\di t}=\sum_{j=1}^r (\beta_{ij}-\alpha_{ij}) k_j  \prod_{l=1}^s c_l(t)^{\alpha_{lj}},\quad i=1,\ldots,s.
\end{align}
For ease of notation, we may  write \eqref{eq:odesystem} as
\begin{align*}
    \frac{\di c}{\di t}=\sum_{j=1}^r \gamma_{ij} \mathcal F_j(\zeta),\quad i=1,\ldots,s,
\end{align*}
where $c=(c_1,\ldots,c_s)$, $k=(k_1,\ldots,k_r)$, $\alpha\in \R^{s\times r}$ with entries $\alpha_{ij}$, $\gamma_{ij}=\beta_{ij}-\alpha_{ij}$ and $\zeta=\zeta(k,\alpha,c)$ such that $\mathcal F_j( \zeta)=k_j  \prod_{l=1}^s c_l^{\alpha_{lj}}$ for $j=1,\ldots,r$.
Systems of the form \eqref{eq:odesystem} do not possess an analytic solution in general. An ODE solver can be used to obtain a numerical solution to \eqref{eq:odesystem} together with an initial condition $c(0)=(c_1(0),\ldots,c_s(0))$.

\subsection{Associated Structural Graph Models}
\label{sec:intro:subsec:graphmodels}

The rule-based epidemiological framework proposed in this article has a further clear advantage, it can make use of several graph theoretic concepts that have been developed for reaction kinetics. This allows for a more smooth transition and interpretation between so-called structural modelling, based static on relations between basic objects, types and rules, and dynamic modelling, based on time series. This would not be achievable by introducing either stochastic processes or dynamical systems from the start of epidemiological modelling. We refer to \cite{Domijan:2008dg} for a discussion of different concepts that can be generalised to stochastic interpretations of rule-based systems.

\section{Essential Models of an Epidemiological Modelling Framework}
\label{sec:standardmodels}

In this section we discuss some simple epidemiological models  and introduce precursor models  consistent with our framework approach from Section \ref{sec:framework}. The precursor models  will only consider some of the rules required to close the models, and will be studied in detail in subsequent publications. The precursor models focus on the mathematical modelling challenges  mentioned in Section~\ref{sec:intro:subsec:challenges}.

\subsection{SIR Model}
\label{sec:standardmodels:subsec:sir}

The SIRD model is the simplest epidemiological model possible. However, for COVID-19 dynamics, it is far too simple. Nevertheless we can introduce general principles that  govern all epidemiological models describing the spread of COVID-19.

We repeat and specify further the general assumptions used to derive 
the stochastic and differential equation formulations of the rules:

\begin{description}
    \item[Mass Action Principle / Transmission]     The propensities of the single rules are directly proportional to the product of the number of individuals in the source involved species. For two species this is equivalent to individuals randomly bumping into each other. In this case we choose a frequency-dependent transmission \cite{Mccallum2001}, with $\frac{1}{N}$ as part of the factor, neglecting the subtraction of deceased individuals. $N$ can be interpreted as fixed size of a reaction volume, where human beings departed because of COVID-19 left irreplaceable space in the society. Or differently interpreted, as a first approximation, the whole population is not shrinking too much due to disease-induced mortality. We are aware that this is, in some cases, particularly when mortality is high, an oversimplication that will have to be addressed in real-life applications in the future. For switching between different transmissions see \cite{Arino2010}.
    \item[Closed Population] We consider neither natural birth or death, nor migration. All deceased individuals died of COVID.
    \item[Exclusivity] All types or subtypes are mutually exclusive, unless induced by the type hierarchy.   
    \item[Homogeneous Types] All individuals of a specific type or subtype are homogeneous regarding to behaviour and disease, and equally affected by the respective rules. 
    \item[Immunity] By recovering an individual acquires complete everlasting immunity and is not contagious any more, unless otherwise stated. 
    \item[Start] We assume that at time $t = 0$ a certain number of individuals are infected / infectious. This becomes especially important if we consider rules with delays.
    \item[Exponential Waiting Times] Waiting times of the rules are exponentially distributed, unless otherwise stated, for general distributions, see \cite{Bog}.
    \item[Time-independent Rates] Each rate is not explicitly time-dependent. In some special cases, e.g. lockdowns, we consider time-dependence in the sense of being constant between given times.
    \item[Independent Execution of Rules] The execution of each rule is independent of the execution of any other rule.
    \item[Differentiability] We consider ODE formulations as limit cases, i.e. as approximations for very large populations.
\end{description}

In contrast to the notation in the previous sections, we use types with  specific interpretations as in Table~\ref{tab:sec:standard:subsec:sir:types} in the following. 
We denote the types in the rules and the number of individuals of respective types in the propensities with the same italic symbols. For the formulations of ODEs and their analysis we consider the associated concentrations and denote them by roman symbols. By this we emphasize the qualitative properties  of the ODE solution are independent of the actual total number $N$ of individuals.

\begin{table}[htp]
\caption{The types of the standard SIR model.}
\label{tab:sec:standard:subsec:sir:types}
\begin{center}
\begin{tabular}{||c  | c||} 
				\hline
				Type & Interpretation  \\ [0.5ex] 
				\hline\hline
				$S$ & Susceptible  \\ 
				\hline
				$I$ & Infectious  \\
				\hline
				$R$ & Recovered  \\
				\hline
				$D$ & Deceased  \\
				\hline
			\end{tabular}
\end{center}
\label{table:sec:standard:subsec:sir-types}
\end{table}%

For reaction constants $\iota,\rho > 0$, the following rules describe type changes:
%
\begin{equation}
\label{eq:sec:standard:subsec:sir}
\begin{alignedat}{5}
	S + I &\xrightarrow{\iota} && 2 I  &\quad\text{[Infections]}\\
	    I &\xrightarrow{\rho}  &&   R  &\quad\text{[Recovery]}\\
\end{alignedat}
\end{equation}
%
Here, the infectious rate $\iota$ controls the rate of spread, i.e.\ $\iota$ represents the probability of
transmitting the disease between a susceptible and an infectious individual. The recovery rate $\rho = 1/\tau$ is determined by the average duration $\tau$ of the infection. Motivated by the COVID-19 pandemic, where numbers of deceased individuals are highly relevant, we introduce a death rate $\delta$:

\begin{equation}
\label{eq:sec:standard:subsec:sird}
\begin{alignedat}{5}
	I &\xrightarrow{\delta} && D &\quad\text{[Death]}\\
\end{alignedat}
\end{equation}

A model consisting of \eqref{eq:sec:standard:subsec:sir} and \eqref{eq:sec:standard:subsec:sird} is also referred to as an SIRD model. We denote the concentrations by $\cS,\cI,\cR$ and the associated number of individuals by $\nS,\nI, \nR$, i.e.\ $\nS=\cS N$, $\nI =\cI N$ and  $\nR = \cR N$.
For the sake of completeness, we summarise the SIRD model for concentrations $\cS,\cI,\cR$ in \eqref{eq:sec:standard:subsec:sir_ODEconcentrations} with propensities \eqref{eq:sec:standard:subsec:sird-SSApropensities} for the associated number of individuals $\nS,\nI, \nR$:

\begin{multicols}{2}
Propensities with numbers:
\begin{equation}
\label{eq:sec:standard:subsec:sird-SSApropensities}
\begin{aligned}
    \kappa_1(\nS, \nI, \nR, \nD) &= \frac{\iota}{N} \nS \nI,\\
    \kappa_2(\nS, \nI, \nR, \nD) &= \rho \nI,\\
    \kappa_3(\nS, \nI, \nR, \nD) &= \delta \nI.
\end{aligned}
\end{equation}
\columnbreak

ODEs for concentrations:
\begin{equation}
\label{eq:sec:standard:subsec:sir_ODEconcentrations}
\begin{aligned}
   	\frac{\di \cS}{\di t} &= - \iota \cS \cI,\\
   	\frac{\di \cI}{\di t} &= \iota \cS \cI -\rho \cI - \delta \cI,\\
   	\frac{\di \cR}{\di t} &= \rho \cI,\\
   	\frac{\di \cD}{\di t} &= \delta \cI.
\end{aligned}
\end{equation}
\end{multicols}

Another possible addition supported by data on COVID-19 is the loss of immunity:

\begin{align}
\label{eq:sec:standard:subsec:sird-loss}
	\begin{split}
	R &\xrightarrow{ \lambda} S \;\; \text{[Loss of Immunity]}\\ 
	\end{split}
\end{align}

Here, $\lambda > 0$ is the rate with which immunity is lost. Loss of immunity is relevant for forecasting herd immunity or the long-time perspectives of vaccination strategies.

\subsubsection{Numerical Simulations}
\label{sec:standard:subsubsec:sird-numericalSimulations}

\begin{figure}[ht]
    \centering
    \begin{subfigure}{0.49\linewidth}
        \centering
        \includegraphics[width=0.8\linewidth]{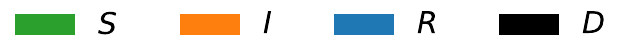}
        \includegraphics[width=\linewidth]{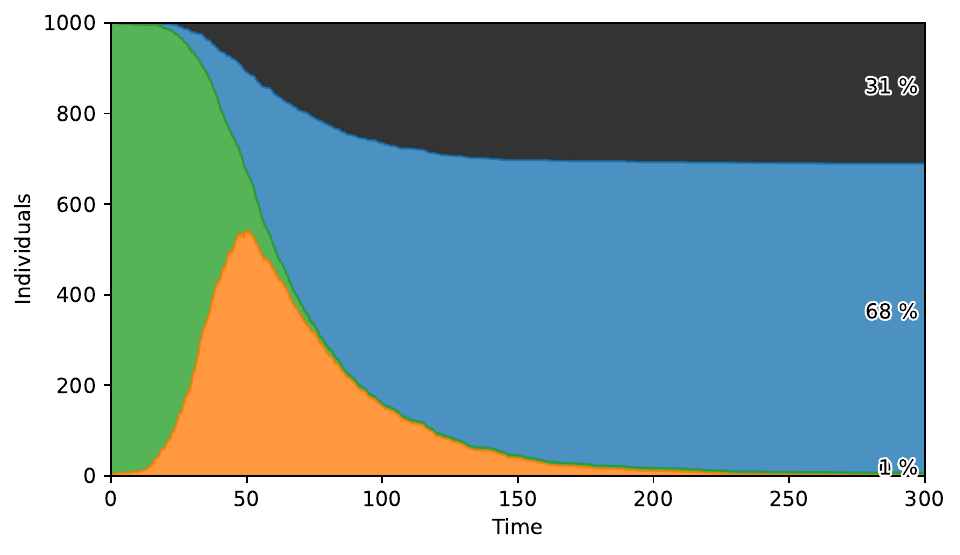}
        \caption{SSA}
        \label{fig:plotSIRD(SSA)}
    \end{subfigure}
    \hfill
    \begin{subfigure}{0.49\linewidth}
        \centering
        \includegraphics[width=0.8\linewidth]{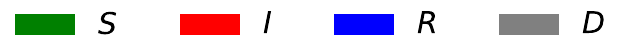}
        \includegraphics[width=\linewidth]{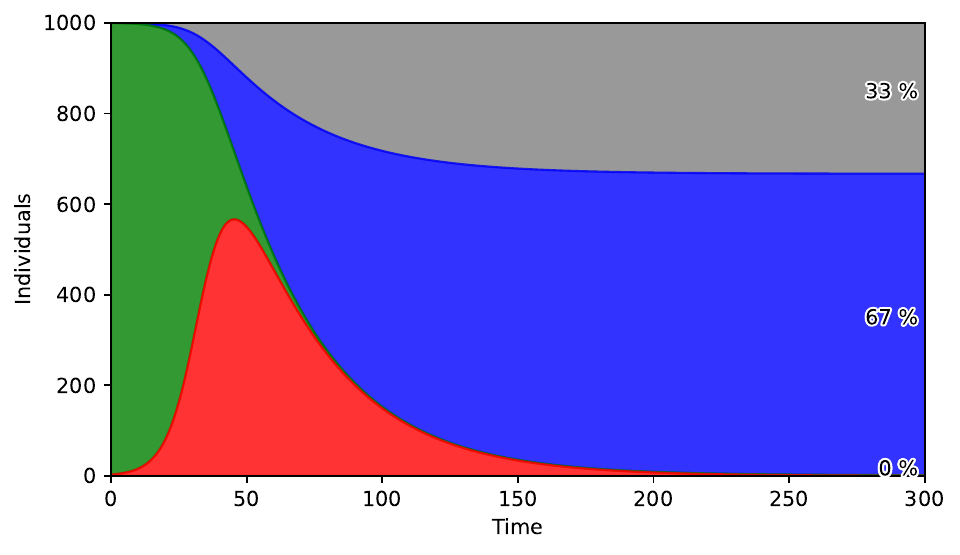}
        \caption{ODE}
        \label{fig:plotSIRD(ODE)}
    \end{subfigure}
    \caption{One realisation of SSA simulation and ODE simulation for SIRD model \eqref{eq:sec:standard:subsec:sir_ODEconcentrations} with basic parameters $N = 1000$, $\iota = \iota^0 = 0.2$, $\rho = \rho^0 = 0.02$, $\delta = \delta^0 = 0.01$, $\nI_0 = \nI_0^0 = 3$.}
    \label{fig:plotSIRD(SSA-ODE)}
\end{figure}

As proof of concept we simulate toy models and try to reproduce features observable in reality. We provide possible interpretations of the results without proof.
All simulations are implemented in python. We focus on SSAs by Gillespie and generalisations, see also \cite{Alonso:2021wm}, but show single characteristic results only for one reproducible random seed. The SSA simulation terminates when time $T = 300$ is reached or when the number of infectious individuals is zero. This results sometimes in an early stop, leaving white space in the figures. 
Also we provide ODE simulations by the standard implementation RK45 of the Runge-Kutta method of order 5 with error control assuming accuracy of fourth-order method in scipy for python.
The ODE system is solved for the fractions, rescaled and stored each integer time value.
For reproducibility and further exploration, we can provide (interactive) Jupyter notebooks for all models.
For the plots we decided to use a stacked representation, highlighting the course for the infected as ground level while simultaneously displaying the other quantities.
We choose similar but distinctive color schemes for the different formulations:
For SSA \includegraphics[height=0.5\baselineskip]{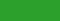} denotes the number of susceptible individuals, \includegraphics[height=0.5\baselineskip]{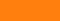} the number of infected/infectious, \includegraphics[height=0.5\baselineskip]{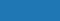} the number of recovered and \includegraphics[height=0.5\baselineskip]{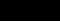} the number of deceased individuals.
Whereas for ODE \includegraphics[height=0.5\baselineskip]{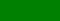} denotes -- here we deviate for easier comparison from fractions -- the number of susceptible individuals, 
\includegraphics[height=0.5\baselineskip]{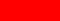} the number of infected/infectious, \includegraphics[height=0.5\baselineskip]{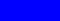} the number of recovered and \includegraphics[height=0.5\baselineskip]{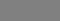} the number of deceased individuals.
When we use types with sub-types, the colors for them differ by brightness. The colors \includegraphics[height=0.5\baselineskip]{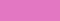} and \includegraphics[height=0.5\baselineskip]{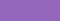} may denote different quantities, like infectivity or vaccination. The parameters are inspired by COVID-19 but not derived explicitly. Therefore we avoid labeling the time unit, but you may think of days. Also the parameters for the rates are exaggerated and adjusted, such that for a total population of $N = 1000$ the different types are clearly recognisable in these small figures. We keep throughout all toy models the basic parameter set $\iota^0 = 0.2$, $\rho^0 = 0.02$ and $\delta^0 = 0.01$ only modifying them according to the extended models. For a coherent compilation see Table~\ref{table:sec:simulation_parameters} in the Appendix.
The initial number of infectious individuals is always $\nI_0^0 = 3$. In Figure \ref{fig:plotSIRD(SSA-ODE)}, numerical results for the SIRD model \eqref{eq:sec:standard:subsec:sir_ODEconcentrations} with basic parameters for a SSA and an ODE solver are shown. After an exponential growth for the number of infectious, their number decrease due to the reduced number of susceptible individuals absorbed by the recovered and deceased. For this single realisation of SSA in Figure~\ref{fig:plotSIRD(SSA)} 3 susceptible individuals survive the epidemic. The ODE simulation in Figure~\ref{fig:plotSIRD(ODE)} reveals a strictly positive share of susceptibles. The SSA and ODE simulations are relatively similar for this parameter setting and chosen random seed.
\subsubsection{Steady states of the SIR Model}\label{sec:steadystatesir}
Let us consider the steady states of the SIR model \eqref{eq:sec:standard:subsec:sir} which can be written as the following system of ODEs:
\begin{align}
\label{eq:sec:standard:subsec:sir_ODE}
\begin{split}
   	\frac{\di  \cS}{\di t} &= - \iota \cS  \cI,\\
   	\frac{\di  \cI}{\di t} &= \iota  \cS  \cI -\rho  \cI, \;\; \\
   	\frac{\di  \cR}{\di t} &= \rho  \cI,
\end{split}
\end{align}
subject to initial conditions $\cS(0)=\cS_0$, $\cI(0)=\cI_0$ and $\cR(0)=\cR_0$ for $\cS_0,\cI_0,\cR_0\in [0,1]$.
The conservation of the total number of individuals implies that $\cS+ \cI+ \cR=1$ at all times $t$.
Hence, the system is in fact two-dimensional and we can consider 
\begin{align*}
   	\frac{\di  \cS}{\di t} &= - \iota \cS \cI, \\
   	\frac{\di  \cR}{\di t} &= \rho \cI.
\end{align*}
We observe that
\begin{align*}
   	\frac{\di  \cS}{\di  \cR}&=-\frac{\iota \cS}{\rho}.
   \end{align*}
Any steady state $(\overline{ \cS},\overline{ \cI},\overline{ \cR})$ satisfies $\overline{ \cI}=0$, implying
\begin{align}\label{eq:stationary_S}
\overline{ \cS}=\cS_0\exp\left(-\frac{\iota\overline{ \cR}}{\rho}\right)={ \cS_0}\exp\left(-\frac{\iota}{\rho}\right)\exp\left(\frac{\iota\overline{ \cS}}{\rho}\right)
\end{align}
for some constant ${ \cS_0}\in [0,1]$ where we used that $\overline{ \cS }+ \overline{ \cR}=1$. 
Here, ${ \cS_0}$ denotes the initial condition of $\overline{ \cS}$, i.e. ${ \cS}(0)={ \cS_0}$,  and implies that $\cR(0)=1-\cS_0$. Note that there exists a unique $\overline{ \cS}\in [0,1]$ since the left- and right-hand side of \eqref{eq:stationary_S} are strictly increasing in $\cS$ for $\cS_0\in(0,1]$ and we have $0<\cS_0\exp(-\tfrac{\iota}{\rho})$ and $1\geq \cS_0$. Hence,  $\overline{ \cR}\in[0,1]$ is also unique and satisfies $\overline{ \cS}+\overline{ \cR}=1$. The disease-free steady state is given by $(\cS,\cI,\cR)=(1,0,0)$. Also note that $\overline{ \cS}$ is always positive for $\cS_0 > 0$, which also holds for $\cS(t)$, as it is monotonically decreasing in $t$ by \eqref{eq:sec:standard:subsec:sir_ODE}.
If the solution $(\cS,\cI,\cR)$ becomes stationary in the long-time limit $t\to \infty$, it converges to the unique steady state $(\overline{ \cS},\overline{ \cI},\overline{ \cR})$. The steady states of the SIRD model can be determined in a similar way, where the disease-free steady state is given by $(\cS,\cI,\cR, \cD)=(1,0,0,0)$.
%
%
\subsubsection{A feature for measuring the transient phase}
\label{sec:standard:subsubsec:sird-transient}
%
If the solution $(\cS,\cI,\cR)$ for given initial data $(\cS_0,\cI_0,\cR_0)$ becomes stationary in the long-time limit $t\to \infty$, it converges to the unique steady state $(\overline{ \cS},\overline{ \cI},\overline{ \cR})$ according to Section \ref{sec:steadystatesir}. To capture the transient phase, we derive the maximum number of infected individuals, following the approach in \cite{Brauer2012}. Due to \eqref{eq:sec:standard:subsec:sir_ODE},  the necessary condition for the maximum of $\cI$ satisfies
\begin{equation*}
    \frac{\di \cI}{\di t} = (\iota \cS - \rho) \cI = 0.
\end{equation*}

Thus,  $\cS = \frac{\rho}{\iota}$ is a critical state and  physically relevant  provided $\rho\leq \iota$ so that $\cS\in [0,1]$ is indeed a concentration. In this case the sufficient condition satisfies $$\frac{\di^2 \cI}{\di t^2} = \iota  (-\iota \cS \cI)  \cI + (\iota \cS - \rho)  \frac{\di \cI}{\di t} = - \rho \iota \cI^2 < 0,$$ implying that $\cI>0$ is maximal when $\cS = \frac{\rho}{\iota}$. 

Dividing the first equation of \eqref{eq:sec:standard:subsec:sir_ODE} by $\cS > 0$  and integrating from $0$ to $t$ yields

\begin{align}
    \label{eq:sec:standard:susbsec:sir_solutionSI}
    \log \cS(t) - \log \cS(0) = - \iota \int_0^t \cI(t') \di t'
    = \frac{\iota}{\rho}  \int_0^t \frac{\di }{\di t'}(\cS(t') + \cI(t')) \di t'= \frac{\iota}{\rho} (\cS(t) + \cI(t) - \cS(0) - \cI(0)).
\end{align}

Here, we used in the second equality that $\frac{\di}{\di t}(\cS + \cI) = - \rho \cI$ which follows from adding the first two equations of
\eqref{eq:sec:standard:subsec:sir_ODE}. 

By inserting $\cS(t) = \frac{\rho}{\iota}$ and solving for $\cI(t)$ we obtain that $\cI(t)$ attains its maximum $\cI_\text{max}$ satisfying

\begin{align*}
    \cI_{max}=\frac{\rho}{\iota}\left(\log \frac{\rho}{\iota} - \log \cS(0) -1\right)+ \cS(0) + \cI(0).
\end{align*}

Note that the transient phase for the SIRD model can be studied in a similar way and requires to replace $\rho$ in the calculations above by $\rho+\delta$.

\subsubsection{$R_0$ and the SIR Model}
\label{sec:standard:subsubsec:sird-R0}

The number of new infections generated by an infected person during their infectious period in a fully susceptible population is an important quantity in classic epidemiology and is referred to as  $R_0$. The parameter $R_0$ characterises an infectious disease and measures the potential of disease expansion at the moment of disease introduction
  in a fully susceptible population. In the absence of disease, the state of the system is called {\it disease-free equilibrium} with $(\cS, \cI, \cR) = (1, 0, 0)$. $R_0$ is related to the stability of the  equilibrium, and also defines a transcritical bifurcation where the disease-free equilibrium changes from being stable,  i.e.\ the disease fails to invade, to being unstable, i.e.\ the disease starts expanding exponentially. When disease transmission is modeled by a system of ODEs, such as in \eqref{eq:sec:standard:subsec:sir_ODE}, the eigenvalues of the linearisation at the disease-free equilibrium are related to $R_0$ in a  straightforward way: $R_0 > 1$ is equivalent to at least one eigenvalue being positive.\\ 

 As a simple example of the direct calculation of $R_0$, we  write the infection rate $\iota$ as a product of the number of contacts an individual has per unit time, denoted by the encounter rate $\beta$, and the probability $p$ of infection given an encounter. This allows us to write $\iota=\beta\,p$ and the SIR model \eqref{eq:sec:standard:subsec:sir_ODE} can be written as
 
\begin{align*}
    \frac{\di \cS}{\di t} &= - { \beta} p \cI \cS\\
    \frac{\di \cI}{\di t} &=   {\bf \beta} p \cI \cS - \rho \cI\\
    \frac{\di \cR}{\di t} &=   \rho \cI.
\end{align*}
The probability $p$ can be further decomposed into
\begin{equation}
\label{eq:SIR_susceptibilityInfectivityFactors}
    p = s i,
\end{equation}
where the susceptibility $s$ measures the risk of getting infected by an encounter, and the infectivity $i$ describes the chance of infection.

At the disease free equilibrium with $\cS = 1$, we can approximate the dynamics of the infected population as
\begin{align*}
    \frac{\di \cI}{\di t} &= \left({\beta} p - \rho\right) \cI 
\end{align*}
with solution 
\begin{align*}
    \label{eq:SIR_solutionEquiI}
    \cI(t)                &= \cI_0 \exp(\left({\beta} p - \rho\right) t)
\end{align*}
for initial data $\cI(0)=\cI_0$.
Hence, the associated eigenvalue of the linearisation is given by $\beta p -\rho$ and the positivity of $\beta p -\rho>0$ is equivalent to 

\begin{equation*}
    R_0   =  \frac{{\beta} p}{\rho}>1,
\end{equation*}

based on the equivalence  mentioned above. Notice that $R_0$ can be regarded as the product of the three factors $\beta$, $p$ and $\tfrac{1}{\rho}$ which correspond to the three fundamental ways by which disease transmission can be controlled: decreasing the number $\beta$ of contacts, decreasing the probability $p$ of infection for every contact (for instance by using  masks) and decreasing  the infectious period (for instance via disease treatment).
This is consistent with the definition of $R_0$ by the largest eigenvalue of the next-generation matrix (NGM) which we denote by $K$ in the following. For techniques to determine $R_0$, see \cite{Diekmann2010}. In our simple case, 
\begin{equation*}
    K = \frac{\iota}{\rho}  = \frac{\beta p}{\rho}
\end{equation*}
is a 1$\times$1-matrix obtained from the infected subsystem of \eqref{eq:sec:standard:subsec:sir_ODE}. Thus,
\begin{equation*}
    R_0 = \frac{\iota}{\rho} =  \frac{\beta p}{\rho}.
\end{equation*}

When including types and rules for deceased individuals in the SIRD model, we derive the basic reproduction number $R_0$ analogously by replacing $\rho$ with total removement rate $\rho + \delta$. Thus we get for the basic parameter setting
\begin{equation}
    \label{eq:SIRD_R0basicSetting}
    R_0^0 = \frac{\iota^0}{\rho^0 + \delta^0} = \frac{0.2}{0.02 + 0.01} = 6 \frac{2}{3}.
\end{equation}

The basic reproduction number $R_0$ not only describes the beginning of epidemics. $R_0$  can also be  used directly for simple cases to describe the fraction of susceptible individuals that survive without any infection. Considering  the limit for $t \rightarrow \infty$ in \eqref{eq:sec:standard:susbsec:sir_solutionSI}, where we set $\cI(0)=0$, $\lim_{t\to\infty}\cI(t) = 0$ and $\cS(0) = 1$,  yields 
\begin{equation}
    \label{eq:finalSizeRelation}
    \lim_{t\to \infty}\cS(t) = \exp \left(- R_0 (1 - \lim_{t\to \infty}\cS(t) \right).
\end{equation}
\eqref{eq:finalSizeRelation} is  called  final size relation  \cite{Brauer2012}, for a meaning-full stochastic derivation, see \cite{Miller2012}. Note that \eqref{eq:finalSizeRelation} is robust to changes within the course of the epidemic.

We can also consider the time-dependent number of new infections an infected individual can generate while infectious and the disease is expanding or established in a population, this is, when susceptible individuals are no longer the whole population. We denote this quantity by $R_{\text{eff}}$ and write $R_{\text{eff}}=S(t)R_0$, i.e.
\begin{equation*}
    R_{\text{eff}}(t) = \frac{\beta p \cS(t)}{\rho}.
\end{equation*}
In this case, an infectious individual encounters fewer susceptible individuals per unit time at an average rate $\beta\,\cS(t)$.\\

One possibility to estimate $R_{\text{eff}}$ from data is to fit a model  and determine the specific value for $R_0$ based on the estimated parameters. A more frequently used method \cite{Cori2013} is to  estimate $R_0$ directly from data, e.g.\ by considering incidences and serial data. The German RKI \cite{anDerHeiden2020} considers the daily reported new cases $I_t^{\text{new}}$ at time $t$ and estimates $R_{\text{eff}}$ from the number of infections within the last four days as
\begin{equation*}
    \widetilde{R}_{\text{eff}}(t) = \frac{\sum_{s = t - 6}^t I_s^{\text{new}}}{\sum_{s = t - 6}^t I_{s-4}^{\text{new}}}.
\end{equation*}

\subsection{Age}
\label{sec:standard:subsec:sir:age}

We make the standard SIR model a bit more realistic by introducing discrete age classes which can vary in vulnerability with respect to the coronavirus. Age was a key factor of individual vulnerability, recognised already at the early stages of the COVID-19 pandemic. \\

\begin{table}[htp]
\caption{The types of the adapted COVID-19 age class structured SIRD model.}
\begin{center}
\begin{tabular}{||c  | r @{\ }l||} 
				\hline
				Types & \multicolumn{2}{c||}{Interpretation}  \\ [0.5ex] 
				\hline\hline
				$S_a, a \in A$ & Susceptible & individuals of age class $a$ \\
				\hline
				$I_a, a \in A$ & Infectious  & individuals of age class $a$ \\
				\hline
				$R_a, a \in A$  & Recovered  & individuals of age class $a$\\
				\hline
				$D_a, a \in A$  & \multicolumn{2}{c||}{Individuals that have died within age class $a$}\\
				\hline
			\end{tabular}
\end{center}
\label{table:sec:standard:subsec:age-sir-types}
\end{table}%

For reaction constants $\iota_{a,a'}, \rho_{a},  \delta_{a} > 0$, $a, a' \in A$, the following rules describe type changes:

\begin{equation}
\label{eq:sec:standard:subsec:age-sird}
\begin{alignedat}{6}
	S_a + I_{a'} &\ccol{\xrightarrow{\iota_{a,a'}}} && I_a + I_{a'}, &\quad a, a' \in A  &\quad\text{[Infections]}\\
	      I_a    &\ccol{\xrightarrow{\rho_{a}}}     && R_a,          &\quad    a  \in A  &\quad\text{[Recovery]}\\
	      I_a    &\ccol{\xrightarrow{\delta_{a}}}   && D_a,          &\quad    a  \in A  &\quad\text{[Death]}\\
\end{alignedat}
\end{equation}

Here, the infectious rates $\iota_{a,a'}$ control the rate of spread, i.e.\ $\iota_{a,a'}$ represents the probability of transmitting the disease between a susceptible of age class $a$ and an infectious individual of age class $a'$. The recovery rates $\rho_a$ and death rates $\delta_a$ are related to the average duration $\tau_a$ of the infection for age class $a$ by $\rho_a + \delta_a = 1/\tau_a$. We assume lower recovery rates for higher age classes, as older individuals may suffer longer from COVID-19, and higher death rates $\delta_a$ for older age classes, as the probability to die is much higher.\\

The NGM for age model (using again Diekmann splitting \cite{Diekmann2010}) is given by:

\begin{equation*}
    K = K_L = %
    \begin{pmatrix}
        p_1 & \ldots & 0 \\
        \vdots & \ddots & \vdots\\
        0   &\ldots & p_{n_a} 
    \end{pmatrix}%
    \begin{pmatrix}
        \iota_{1,1} & \ldots & \iota_{1, n_a} \\
        \vdots & \ddots & \vdots\\
        \iota_{n_a, 1}   &\ldots & \iota_{n_a, n_a}
    \end{pmatrix}%
    \begin{pmatrix}
        \frac{1}{\rho_1 + \delta_1} & \ldots & 0 \\
        \vdots & \ddots & \vdots\\
        0   &\ldots & \frac{1}{\rho_{n_a} + \delta_{n_a}} 
    \end{pmatrix},
\end{equation*}
where $p_a$ is the fraction of all individuals in age class $a$.

Besides recovery and death, age creates heterogeneity by structuring these groups by different contact rates and infectivity. Thus the infection rate matrix becomes dependent on susceptibility $s_a$, infectivity $i_a$ and contacts $C_{a, a'}$ of the respective age classes $a, a'$

\begin{equation}
\label{eq:sec:standard:subsec:age-sird:matrix_infectivity}
    \iota_{a, a'} = q \cdot s_a \cdot i_{a'} \cdot C_{a, a'},
\end{equation}

where $q$ is a calibration factor accounting for parameters on the same time scale. Note that by different values for susceptibility and infectivity the infection matrix is not necessarily symmetric.
For these simple models we assume constant contact matrices, which of course is not very realistic as behaviour induced by the epidemic changes contacts.

\begin{figure}[ht]
    \centering
    \includegraphics[height=3.em]{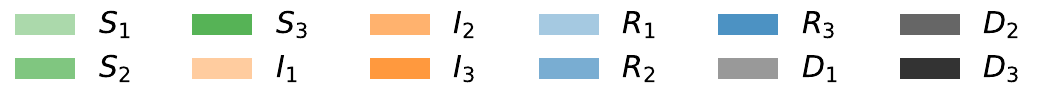}
    \begin{subfigure}{0.49\linewidth}
        \centering
        \includegraphics[width=\linewidth]{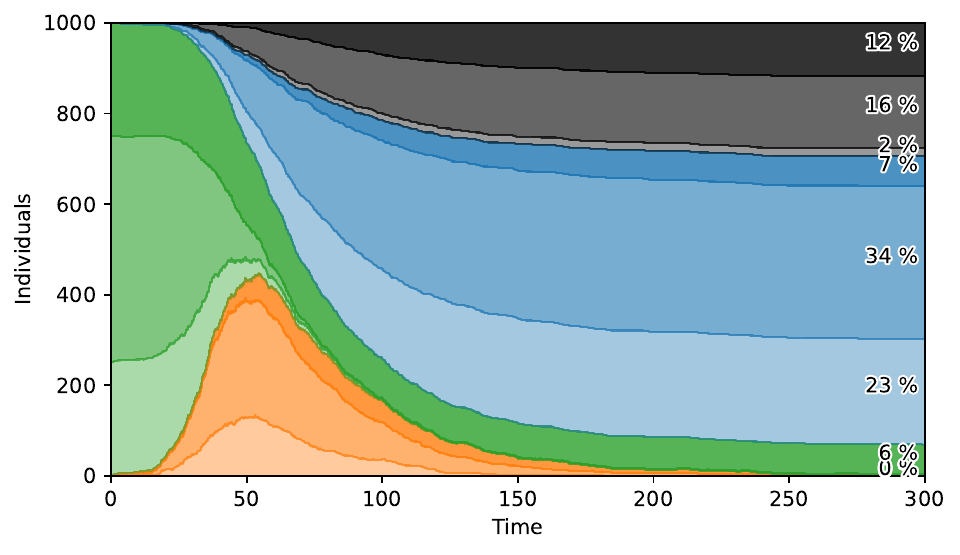}
        \caption{contact $100\%$, $R_0 \approx 6.66$ \newline}
        \label{fig:plotAge(SSA)_Contact}
    \end{subfigure}
    \hfill
    \begin{subfigure}{0.49\linewidth}
        \centering
        \includegraphics[width=\linewidth]{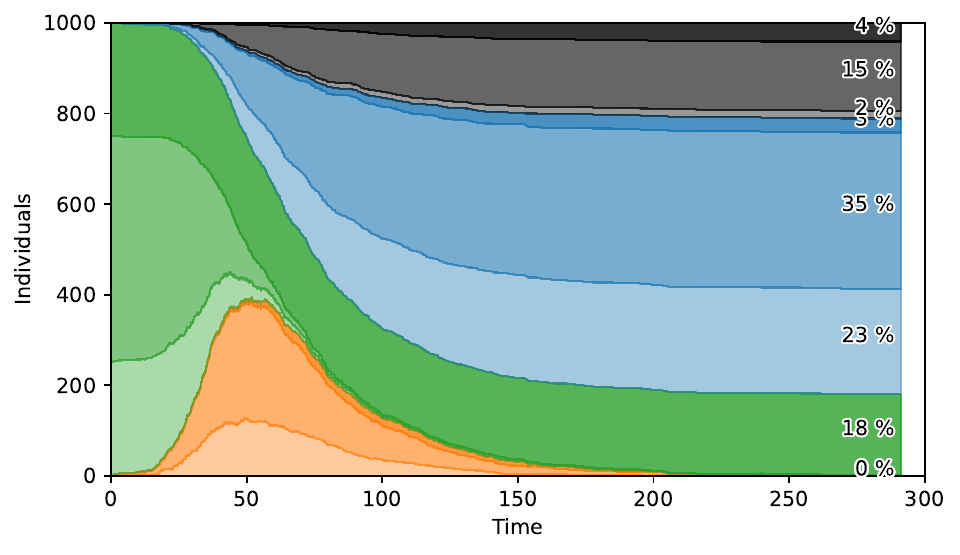}
        \caption{contact reduction of $80\%$ with oldest age class, $R_0 \approx 6.59$}
        \label{fig:plotAge(SSA)_ContactRed}
    \end{subfigure}
    
    \caption{SSA simulations for SIRD model with age including contact matrices. $\rho = \rho^0 (1.5, 1.0, 0.5)$, $\delta = \delta^0 (0.25, 1.0, 2.)$, $p = (0.25, 0.5, 0.25)$.}
    \label{fig:plotAge(SSA)}
\end{figure}

For the simulations in Figure~\ref{fig:plotAge(SSA)} we assumed for reasons of simplification constant susceptibility and infectivity but different parameters for recovery and death.
We choose the contact matrix $C$ based on condensed and symmetrised results by \cite{Wallinga2006}, see also \cite{Mossong2008,Klepac2020,Prem2021}

\begin{equation*}
    C = %
    \begin{pmatrix}
        76 & 38 &  5\\
        38 & 67 & 13\\
         5 & 13 & 9
    \end{pmatrix}
\end{equation*}

and the calibration factor $q$ such that the basis reproduction number in Figure~\ref{fig:plotAge(SSA)_Contact} equals $R_0$ of the basis case \eqref{eq:SIRD_R0basicSetting}, see Figure~\ref{fig:plotSIRD(SSA-ODE)}. Here we observe the hidden heterogeneity reducing the maximum number of infected, but also less recovery and larger fatality for the oldest age group increasing their death fraction to almost half of the beginning.
In Figure~\ref{fig:plotAge(SSA)_ContactRed} we kept the calibration and all other parameters unchanged but modeling saving vulnerable groups, here the elderly, by young and middle aged reducing contacts with them by $80\%$. This results in a marginally smaller value for $R_0$ but in a significantly smaller number for fatalities in the oldest group.

\subsection{Global Information and State Dependent Rates}
\label{sec:standard:subsec:sir:global}

One aspect making pandemics in the 21st century different from pandemics in previous centuries is the  availability of  data  on the disease, from local to even global levels. We  discuss a simple $SIRD$-model which incorporates this data with the help of global information functional responses, see Section \ref{sec:intro:subsec:dynamics:sub:mc}. For the classification, we use Table \ref{table:sec:standard:subsec:sir-types}. The rules now incorporate functional responses:

\begin{equation}
\label{eq:sec:standard:subsec:sir-fr}
\begin{alignedat}{3}
	S + I &\ccol{\xrightarrow{\iota(S, I, R, D; \mu)}} && I + I  &&\quad\text{[Infection]}\\
	    I &\ccol{\xrightarrow{\rho}}   && R &&\quad\text{[Recovery]}\\
	    I &\ccol{\xrightarrow{\delta}} && D &&\quad\text{[Death]}\\
\end{alignedat}
\end{equation}

Here, we suppose that the infection rate $\iota$ may depend on $S,I,R,D$ as well as additional parameters summarised in the variable $\mu$. We assume that $\iota$ is monotonically increasing in $S$ and $R$,  modelling that if the population registers many unharmed (but susceptible) and recovered individuals,  the population in general becomes more careless.  We assume that $\iota$ is monotonically decreasing  in $I$ and $D$, modelling that if the population detects more infected, or even dead individuals, then the population in general becomes more careful in behavioural terms.
In this simple model  the infection rate in the form of a functional response reduces to 

\begin{equation*}
    \iota(\nI; \mu) = \frac{\iota_0}{1 + \mu \frac{\nI}{N}} \quad \text{(used in propensities),} \quad \iota(\cI; \mu) = \frac{\iota_0}{1 + \mu \cI} \quad \text{(used in ODEs),}
\end{equation*}

where $\mu$ is a non-negative parameter that models the severeness of contact reductions. Note that this function works as a self-regulator:  for increasing numbers $\nI$, individuals may become more cautious and follow social distancing rules which may lead to a decreasing infection rate $\iota$ and a decrease of the number of newly infected individuals, while at a later point in time the infections may rise again. This results in the following propensities for numbers $\cS,\cI,\cR, \cD$ of individuals and the associated concentrations $\nS,\nI,\nR, \nD$:

\begin{multicols}{2}
Propensities with numbers:
\begin{equation}
\label{eq:sec:standard:subsec:sird-fr-SSApropensities}
\begin{aligned}
    \kappa_{1}(\nS, \nI, \nR, \nD) &= \frac{\iota(\nS, \nI, \nR, \nD; \mu)}{N} \nS \nI,\\
    \kappa_{2}(\nS, \nI, \nR, \nD) &= \rho \nI,\\
    \kappa_{3}(\nS, \nI, \nR, \nD) &= \delta \nI.\\
\end{aligned}
\end{equation}
\columnbreak
ODEs with fractions:
\begin{equation}
\label{eq:sec:standard:subsec:sir-fr_ODEconcentrations}
\begin{aligned}
   	\frac{\di \cS}{\di t} &= -\iota(\cS, \cI, \cR, \cD; \mu) \cS \cI,\\
   	\frac{\di \cI}{\di t} &=  \iota(\cS, \cI, \cR, \cD; \mu) \cS \cI - (\rho + \delta) \cI,\\
   	\frac{\di \cR}{\di t} &= \rho \cI,\\
   	\frac{\di \cD}{\di t} &= \delta \cI.
\end{aligned}
\end{equation}
\end{multicols}

Note that  (\ref{eq:sec:standard:subsec:sir-fr_ODEconcentrations}) always conserves the concentration of all individuals, i.e.

\begin{align}
\label{eq:conserved_number}
\cS(t)+\cI(t)+\cR(t)+\cD(t)=1
\end{align}

for all times $t\geq 0$. We assume that infection rate $\iota(\cI,\cR,\cD;\mu)$ is a positive, differentiable function of $\cI$, $\cR$ and $\cD$ which increases with respect to $\cR$ and decreases with respect to $\cI$ and $\cD$:

\begin{equation}
\label{eq:conditions}
\begin{aligned}
    &\frac{\partial \iota(\cI,\cR,\cD;\mu)}{\partial \cR}>0\mbox{ for all $(\cR,\cI,\cD)\in [0,1]^3$ and $\mu\in\R^\ell$},\\
    &\frac{\partial \iota(\cI,\cR,\cD;\mu)}{\partial \cI}<0\mbox{ for all $(\cR,\cI,\cD)\in [0,1]^3$ and $\mu\in\R^\ell$},\\
    &\frac{\partial \iota(\cI,\cR,\cD;\mu)}{\partial \cD}<0\mbox{ for all $(\cR,\cI,\cD)\in [0,1]^3$ and $\mu\in\R^\ell$}.
\end{aligned}
\end{equation}

The conditions (\ref{eq:conditions}) describe the effects of available information:

\begin{itemize}
\item the more \emph{recovered} individuals exist, the more there is an increase of interactions among individuals and thus increasing infectivity. 
\item The more \emph{infections} the more appears a tendency to decrease of interactions among individuals and thus decreasing infectivity.
\item The more \emph{deaths} the more appears a tendency to decrease of interactions among individuals and thus decreasing infectivity.
\end{itemize}

\begin{figure}[ht]
    \centering    
    \begin{subfigure}{0.49\linewidth}
        \centering
        \includegraphics[height=1.5em]{plotsRBSEa1/legends/SIRD_SSA-h.pdf}
        
        \includegraphics[width=\linewidth]{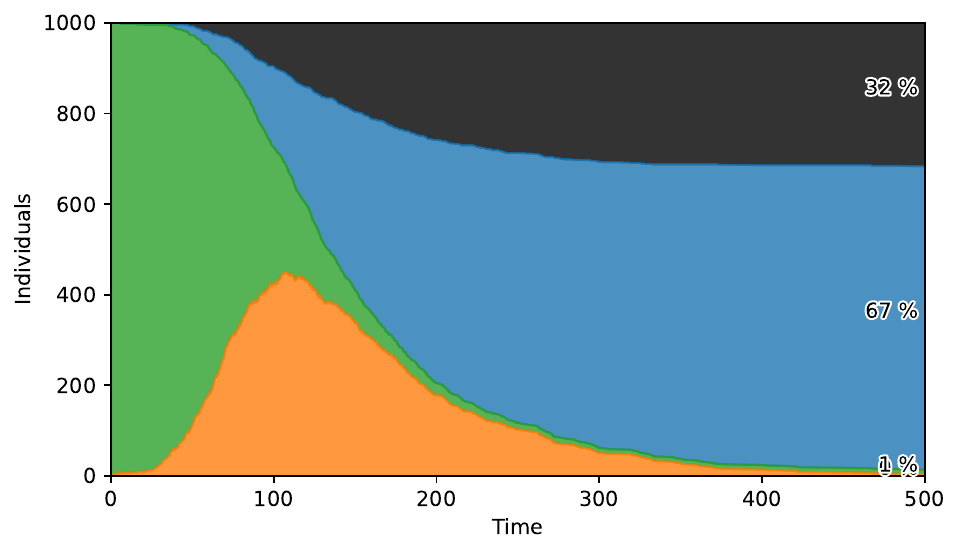}
        \caption{SSA, $N=1000$, $\mu = 1$}
        \label{fig:plotSIRDresp(SSA)_lambda1}
    \end{subfigure}
    \hfill
    \begin{subfigure}{0.49\linewidth}
        \centering
        \includegraphics[height=1.5em]{plotsRBSEa1/legends/SIRD_SSA-h.pdf}
        
        \includegraphics[width=\linewidth]{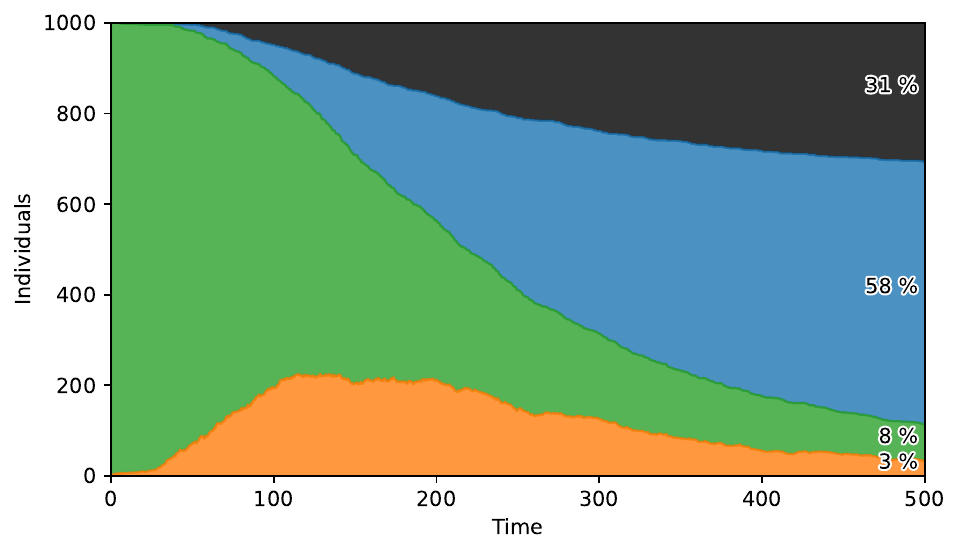}
        \caption{SSA, $N=1000$, $\mu = 10$}
        \label{fig:plotSIRDres(SSA)_lambda10}
    \end{subfigure}
    \caption{SSA simulations for SIRD model with functional response, for base case parameters and simple saturation $\iota(\nI; \mu) = \frac{\iota^0}{1 + \mu \frac{\nI}{N}}$ for different $\mu$.}
    \label{fig:plotSIRDres(SSA)}
\end{figure}

\begin{figure}[ht]
    \centering
    \begin{subfigure}{0.49\linewidth}
        \centering
        \includegraphics[width=0.9\linewidth]{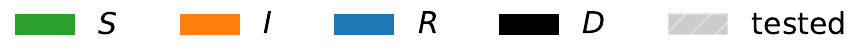}
        \includegraphics[width=\linewidth]{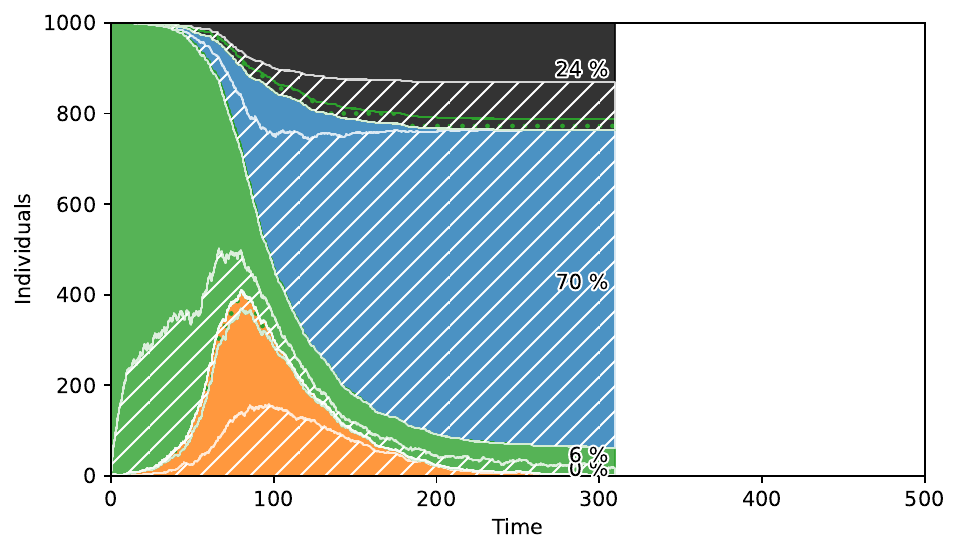}
        \caption{SSA, $N=1000$, $\mu = 1$}
        \label{fig:plotSIRDtresp(SSA)_lambda1}
    \end{subfigure}
    \hfill
    \begin{subfigure}{0.49\linewidth}
        \centering
        \includegraphics[width=0.9\linewidth]{plotsRBSEa1/legends/SIRDt_SSA-h.pdf}
        \includegraphics[width=\linewidth]{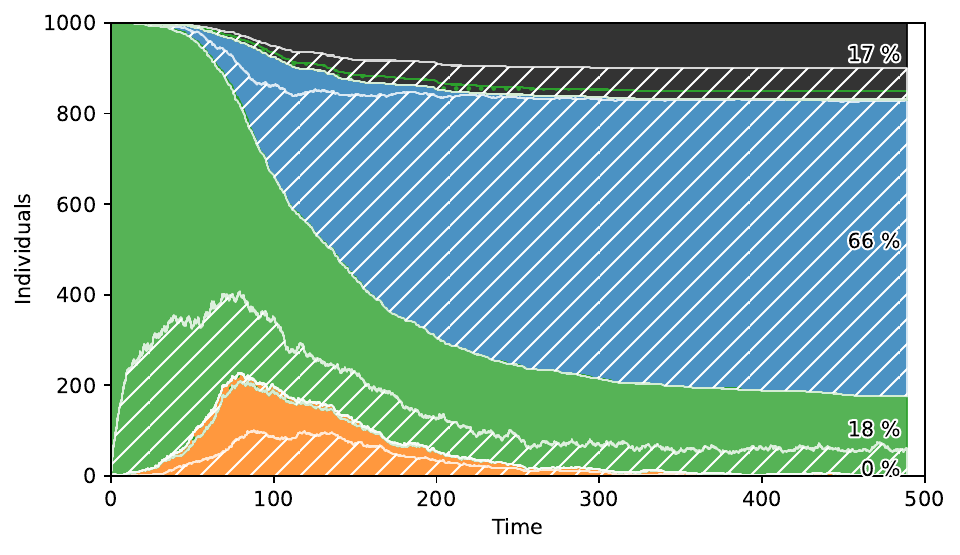}
        \caption{SSA, $N=1000$, $\mu = 10$}
        \label{fig:plotSIRDtres(SSA)_lambda10}
    \end{subfigure}
    \caption{SSA simulations for SIRD model with testing (minor testing $\theta$, higher loss $\lambda$) and functional response, simple saturation $\iota(\nI; \mu) = \frac{\iota^0}{1 + \mu \frac{\nI_1}{N}}$  for different response $\mu \in \R$, loss of testing information rate $\lambda = 0.1$.}
    \label{fig:plotSIRDtres(SSA)}
\end{figure}

For simulations by SSA for $\mu=1$ and $\mu=10$, see Figure~\ref{fig:plotSIRDres(SSA)}. Note that here the maximum time is set to $T = 500$.
Even more realistically, we can model the global information based on the number of tested infectious individuals $\nI_1$, bringing in an uncertainty aspect. Numerical results for such a model are shown in Figure~\ref{fig:plotSIRDtres(SSA)}. Comparing Figure~\ref{fig:plotSIRDtres(SSA)_lambda10} to Figure~\ref{fig:plotSIRD(SSA)} it is possible to observe that testing and global information lead to a more effective flattening the curve of the infected cases and lowering the death toll. The contribution of testing can be seen by considering Figure~\ref{fig:plotSIRDres(SSA)_lambda10} where the death exceeds the level of Figure~\ref{fig:plotSIRDtres(SSA)_lambda10}.

Note that the shape of the hashed parts of Figure~\ref{fig:plotSIRDtres(SSA)} look similar to Figure~\ref{fig:plotSIRDres(SSA)}. The number of unrecorded cases are obviously greater for global information dependent on tested cases, but can be controlled with appropriate behaviour $\mu$.

\subsubsection{Steady states}
\label{sec:standard:subsec:sir:global:analysis}

In this subsection we want to explore a simple model aiming to study what can be called \emph{global information}, namely the fact that the knowledge of $I$, $R$ and $D$ may influence the behaviour of many individuals who may try to adjust their actions (e.g.\ social distancing, precaution measures, etc), in order to minimise their risk of infection. To this end we require an infection rate depending upon infectious cases $I$, recovered cases $R$ and dead cases $D$.  In particular we study the dependence of the maximum of the infection on the form infection rate. We make the following assumptions:

 \begin{itemize}
 \item the information is  available uniformly in the population,
 \item the information  affects the rates instantaneously.
 \end{itemize}
Consider a SIR model with $(\cS,\cI,\cR,\cD)\in [0,1]^4$ with an infection rate $\iota(\cI,\cR,\cD;\mu)$ as a smooth function of $\cI$, $\cR$, $\cD$ and a set of parameters denoted by a vector $\mu\in \R^\ell$ for some $\ell\geq 1$. Following the notation in the previous section we consider the infection rate $\iota(\cI,\cR,\cD;\mu)$ which results in a SIRD model with state dependent infection rate: 

\begin{align}
\label{eq:SIR_i}
\begin{split}
&	\frac{\di \cS}{\di t} = -\iota(\cI,\cR,\cD;\mu)\,\cS\,\cI,\\
&	\frac{\di \cI}{\di t} = \iota(\cI,\cR,\cD;\mu)\,\cS\,\cI-\rho\,\cI-\delta\,\cI,\\
&	\frac{\di \cR}{\di t} = \rho\,\cI,\\
&   \frac{\di \cD}{\di t} = \delta\,\cI.
\end{split}
\end{align}

The steady states $(\bar \cS,\bar \cI,\bar \cR,\bar \cD)$ are characterised by the  conditions

\begin{align*}
\overline{\cI}=0, \qquad
\overline{\cS}+\overline{\cR}+\overline{\cD}=1,
\end{align*}

implying that all steady states are \emph{infectious free} states. Under the above assumptions on $\iota$, \eqref{eq:SIR_i} has a unique solution $\cS(t),\cI(t),\cR(t),\cD(t)$ which depends continuously on $\mu$. By
 (\ref{eq:SIR_i}) there exists a manifold $\Sigma$ of states satisfying $\frac{\di \cI}{\di t}=0$ and given by
 
\begin{align*}
\iota(\cI,\cR,\cD;\mu)\,\cS-\rho-\delta=0, \quad
\cS+\cI+\cR+\cD=1.
\end{align*}
Equivalently, $\Sigma$ can be characterised by the single equation
\begin{align}
\label{eq:Sigma}
\iota(\cI,\cR,\cD;\mu)\,(1-\cI-\cR-\cD)-\rho-\delta=0.
\end{align}

Under appropriate assumptions on $\iota$, e.g.\ $\rho+\delta \leq \iota(\cI,\cR,\cD;\mu) $ for all $\cI,\cR,\cD$ as in Section \ref{sec:standard:subsubsec:sird-transient}, a solution to \eqref{eq:Sigma} is guaranteed. Using \eqref{eq:SIR_i}, we obtain

\begin{align*}
\frac{\di^2 \cI}{\di t^2}&=\frac{\di \cI}{\di t}(\iota(\cI,\cR,\cD;\mu)\,\cS-\rho-\delta)+\cI\left(\frac{\di \cS}{\di t}\,\iota(\cI,\cR,\cD;\mu)+\cS\frac{\partial \iota(\cI,\cR,\cD;\mu)}{\partial \cI}\frac{\di \cI}{\di t}\right.\\&\quad\left.+\cS\frac{\partial \iota(\cI,\cR,\cD;\mu)}{\partial \cR}\frac{\di \cR}{\di t}+\cS\frac{\partial \iota(\cI,\cR,\cD;\mu)}{\partial \cD}\frac{\di \cD}{\di t}\right)
\end{align*}

which reduces to 

\begin{align}
\label{eq:time_der2_I}
\frac{\di^2 \cI}{\di t^2}&= -\cI^2 \cS \iota^2(\cI,\cR,\cD;\mu) +\cI \cS \left(\rho \frac{\partial \iota(\cI,\cR,\cD;\mu)}{\partial \cR}+\delta \frac{\partial \iota(\cI,\cR,\cD;\mu)}{\partial \cD}\right)
\end{align}

on $\Sigma$ since $\frac{\di \cI}{\di t}=0$ on $\Sigma$. In general the convexity of $\cI(t)$ on $\Sigma$ depends on the derivatives of the infection rate $\iota$ with respect to $\cR$ and $\cD$. In the special case when $\iota(\cI,\cR,\cD;\mu)=\iota(\cI;\mu)$,  the second time derivative (\ref{eq:time_der2_I}) reduces to 

\begin{align*}
\frac{\di^2 \cI}{\di t^2}=-\cI^2 \cS \iota^2(\cI,\cR,\cD;\mu)<0
\end{align*}

which implies that $\cI(t)$ attains a maximum satisfying \eqref{eq:Sigma}. If $\iota(\cI;\mu)$ is a  monotonically decreasing, smooth function satisfying

\begin{align*}
\lim_{|\mu|\rightarrow+\infty}\iota(\cI;\mu)=0\mbox{ and }
\lim_{|\mu|\rightarrow+\infty}\frac{\partial \iota(\cI;\mu)}{\partial \cI}=0
\end{align*}
then the maximum of $\cI(t)$ flattens as $|\mu|=\max\{\mu_1,...,\mu_\ell\}$ becomes larger. This is consistent with the numerical experiment in Figure \ref{fig:plotSIRDtres(SSA)} illustrating that  $\cI(t)$ tends to be flattened for larger values of the scalar $\mu$. 

For $\mu\in \R$ and the simple saturation $\iota(\cI; \mu) = \frac{\iota_0}{1 + \mu \cI}$ as functional response, we are able to derive this dependency of the flattening, measured by $\cI_{max}$, explicitly. Dividing the second equation of \eqref{eq:sec:standard:subsec:sir_ODE} by the first one, we can analyse the system by focusing on the relation  between $\cI$ and $\cS$
\begin{equation*}
    \frac{\di \cI}{\di \cS} = -1 + \frac{\rho+\delta}{\iota(\cI; \mu) \cS} = -1 + \frac{(\rho+\delta) (\iota_0 + \mu \cI)}{\cS}.
\end{equation*}
This ODE admits for $\mu > 0$ a solution using variation of constants and with initial condition $\cI(S_0) = \cI_0$.

\begin{figure}[ht]
    \centering
    \begin{subfigure}{0.4\linewidth}
        \centering
        \includegraphics[height=0.9\linewidth]{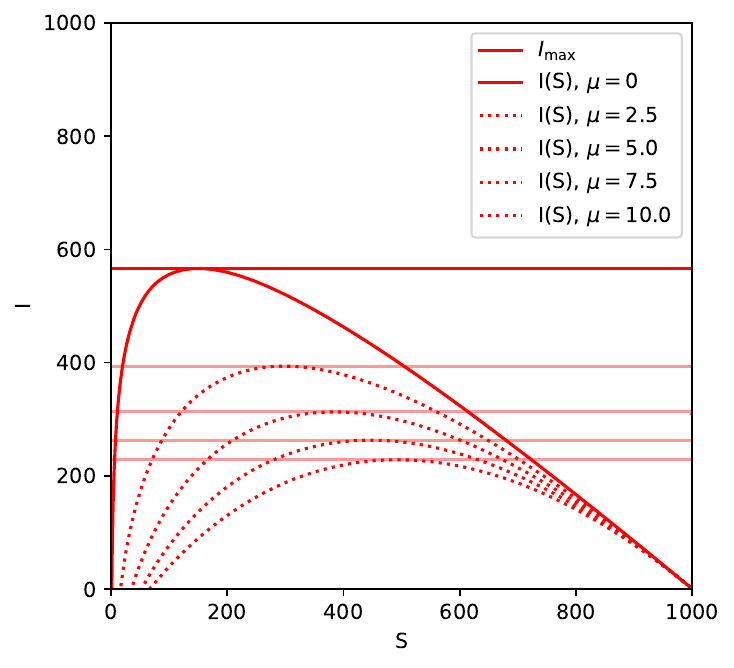}
        \caption{$\nI$ for different parameters $\mu$}
        \label{fig:plotSIRDresp(ODE)-I}
    \end{subfigure}
    \begin{subfigure}{0.4\linewidth}
        \centering
        \includegraphics[height=0.9\linewidth]{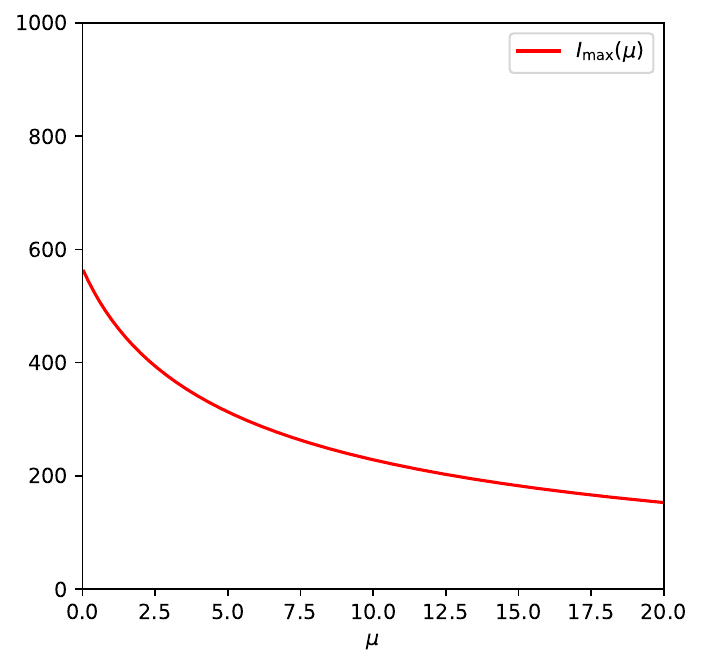}
        \caption{$\nI_{\max}$ of $\mu$}
        \label{fig:plotSIRDres(ODE)-ImaxVSlambda}
    \end{subfigure}
    \caption{ODE simulations for SIRD model with simple saturation response $\iota(\cI; \mu) = \frac{\iota_0}{1 + \mu \cI}$, $\iota_0 = \iota^0 = 0.2$, $\rho = \rho^0 = 0.02$, $\delta = \delta^0 = 0.01$ for different $\mu$.}
    \label{fig:plotImax(ODE)}
\end{figure}

Thus we obtain for $\cI_{max}(\mu)$
\begin{align*}
    \cI_{\max}(\mu) &= \left( \frac{\cS^\ast}{\cS_0} \right)^{(\rho+\delta) \mu} \left( \cI_0 - \frac{\cS_0}{(\rho+\delta) \mu - 1} + \frac{\iota_0}{\mu}\right) + \frac{\cS^\ast}{(\rho+\delta) \mu - 1} - \frac{\iota_0}{\mu},\\
    \text{where } \cS^\ast &= \left( \frac{\cS_0^{(\rho+\delta) \mu}}{\cS_0 - ((\rho+\delta) \mu - 1) \cI_0 - \frac{\iota_0}{\mu}} \right)^{(\rho+\delta) \mu -1}.
\end{align*}
This relation is shown in Figure~\ref{fig:plotImax(ODE)}.

\section{Discussion And Outlook}
\label{sec:discussion}

We have proposed an epidemiological modeling framework for the COVID-19 dynamics, which introduces a systematic and mathematical consistent way to model all aspects of the COVID-19 pandemic, dependent on the questions being asked, policies investigated, and - not covered in this paper, but see Section \ref{sec:intro:subsec:framework} and future work - the \emph{available data}. The framework would have been empty if we would not have introduced model outlines, in some cases complete models, furthermore code and simulations, for all aspects of the COVID-19 pandemic. We therefore believe we have established a new milestone on which future IT-based expert systems can help humanity to manage global pandemics more efficiently.

\subsection{The Prognosis Problem of Mathematical Models}
\label{sec:discussion:subsec:prognosis}

We would like to stress that if mathematical modelling exists in order to help managing a pandemic effectively, the predictive power of models, or family of models, as is the approach in this article, needs to be tested and finally established. Mathematical modelling might also just exist to focus minds of experts, and establish a discussion. Although we support this idea, a restriction on this limited focus would not bring mathematical modelling towards its true capability. With the advance of modern methods of applied mathematics and  computational power we should and also can aim for much more. Unfortunately we could not identify true proven predictive power in \emph{any} of the current mathematical models used during the pandemic. Some of the models might actually have prognostic power, but they rely more on the reputation of work groups or institutions rather than objective criteria. We call this current lack of predictive or prognostic power the \emph{fundamental prognosis problem}, which is intrinsically related to Bayesian model selection, see Section \ref{sec:intro:subsec:framework}. The reasons for the fundamental prognosis problem are manifold:

\begin{description}
\item[Complexity] During the time course of an epidemic the background conditions change such that they are not covered by the current types, i.e. variables of a model. One has to realise that mathematical models are compact descriptions of reality, and therefore neglect circumstances which at a moment, or better a certain time frame, do not significantly contribute to the measured data. However, in a complex system such as an epidemic, the underlying mechanisms might change, therefore the number of variables needed to describe the system might increase.
\item[Model Selection] It is clear that single models, whatever their complexity, are parametrised with a certain data set from the past. Therefore the fundamental assumption is again that the conditions giving parameters their numerical value, or parameter range, should not have significantly changed when a prognosis is attempted.
\item[Model Comparison] In epidemiology, unlike for example in climate modelling, models are not regularly compared with each other to establish strength and weaknesses of prognostic power in certain situations. This would need the international and interdisciplinary cooperation of epidemiological modellers. There are also no established criteria with which prognostic power is measured. Such a discussion is long overdue.
\item[Deterministic versus Stochastic Modeling] A large misunderstanding by some experts and the general public is the lack of understanding of the intrinsic stochastic nature of an epidemic. Deterministic models are good enough in some circumstances, and therefore reliable, but only if the law of large numbers can be applied to all aspects of the model situation. In this case one can establish a sensitivity analysis, showing which parts of the model react most significantly if data are not reliable, or the underlying conditions change only slightly. However, in stochastic modeling the idea of a definitive time course has to be abandoned. It is replaced by the idea of an ensemble of time courses, which all can happen with certain probabilities. Prediction in a stochastic world therefore is more difficult, and has a very different meaning.
\end{description}

\subsection{The Data}
\label{sec:discussion:subsec:data}

We have not covered in this article any methods and applications of statistics and data analysis in general, in particular we have not parametrised our models with real world data. This very important task, a highly complex topic in itself, is left to future considerations and publications. Here we give just some general observation relevant for the presented modelling framework. Most importantly, the COVID-19 pandemic has brought scientific data reading literally to the forefront of billons of people, a highly relevant development for science in general. Reading and listening to the latest infection numbers has become a part of daily routine by billions of people. Of course, like for mathematical modelling discussed in the previous Section \ref{sec:discussion:subsec:prognosis}, there are problems with standardisation of data as well, see \cite{Gardner:2021tv}. The internet and modern IT-infrastructure have made data collection on a massive scale nevertheless feasible. One of the main institutions showing how this can be done is the Johns Hopkins University with its Coronavirus Ressource Center, \cite{JHUniversity:2021vb}. But nearly all national or international organisations related to epidemiology, also newspapers and other institutions, have collected relevant data, see the references in \cite{Callaghan:2020wk}, or the WHO \cite{who2021}. The speed of data collection and display has generally improved \cite{Dong:2020te}. Therefore, there should be plenty of scope to connect in a next step mathematical models and epidemiological data in a more continuous fashion as well. \\

It is important to note that a reliable epidemic model or expert system will have to incorporate a multitude of different data. For example, in order to learn from the effectiveness of policy measures, one needs a clear understanding of the nature of policies, for example lockdowns, and their duration. A good example of available data bases is the Oxford policy database \cite{Mahdi:2021vl}.

\subsection{Future Expert Systems}
\label{sec:discussion:subsec:expert-systems}

The two previous Sections \ref{sec:discussion:subsec:prognosis} and \ref{sec:discussion:subsec:data} on mathematical models and data lead naturally to the question how these two streams of research can work together more effectively in future? A natural environment where these different mathematical and empirical research can converge can be expressed as an idea of a novel type of expert system, based on modern mathematical and computational methods, IT-developments and empirical science. We believe the following topics are the most important ones in a long list of choices and properties of such a novel expert system:

\begin{description}
\item[Rule-Based Systems] Rule-based systems as exclusively used in this article have a huge advantage: they are easy to interpret once some basic intuition for rules can be acquired. This experience can also be acquired by non-mathematicians. Moreover, rules can be easily stored in computers. This should enable a way to use machine learning and artificial intelligence in several ways to aid successful prognosis of a pandemic time course, described below. As long as rule-based systems are simple enough, they can be extremely simply communicated to non-scientists. It has become more more evident, especially during the pandemic, that science in general needs to communicate better to be understood by the public.

\item[Machine Learning, AI] Future expert systems will rely on both mathematical modelling and machine learning. We use here the term machine learning in a general sense, the term includes also artificial intelligence. Artificial intelligence methods have made rapid progress in the last years, especially deep learning methods represent a leap forward in unsupervised automatic learning from data \cite{Russell:2021tz}. A typical application to the Covid-19 pandemic is given by \cite{Shamout:2021uy}. That mathematical models and AI methods should be developed side by side is no contradiction. For example, the claim that deep learning alone could substitute mathematical modelling is far-fetched and dangerous. Mathematical modelling is hypothesis building of underlying mechanisms governing a process, in our case a pandemic process. In order to be able to do this, there needs to be some time invariance of these acting mechanisms. But as long as the time invariance is guaranteed, mechanisms are ideally identified by the mathematical model, and therefore policies can be put in place to manipulate them in a desired direction. Note that this is not the case if one relies on artificial intelligence approaches only. Artificial intelligence cannot identify mechanisms, \emph{ it is simply not designed for this purpose}. \\ 

However, machine learning and AI are needed in other ways, mostly for data integration into mathematical models, we discuss this as a next topic below. Another, perhaps less obvious help by machine learning techniques are Bayesesian model selection algorithms. 
Without a model framework where mathematical models are systematically able to be altered according to empirical evidence, mathematical models will not help to understand epidemics, the prognosis problem, see Section \ref{sec:discussion:subsec:prognosis}, will not be overcome. There are other uses for machine learning as well. For example, one can contemplate how non-mathematical experts formulate their insight into the pandemic by \emph{ verbal descriptions}. By machine learning this verbal description, in addition with mathematical consistency checks, is translated into a rule-based system. It could be a very fruitful way how a future expert system interacts with different experts from different fields, expressing their opinions and subsequently validating and investigating such believe systems with the help of empirical data.

\item[Data Incorporation] Machine learning and AI are urgently needed to integrate real world data into future expert systems based on mathematical models. As more and more data are becoming available online, or at least in a machine readable form. All mathematical models need to be parametrised, and the more data are available for that purpose, the better. There are some promising new approaches for model parameter estimation that should be very suitable for epidemiological modelling problems, like SINDy \cite{Champion:2019vq},  \cite{Kaheman:2020wc}. An application to epidemiology is reference \cite{Horrocks:2020wr}. In  emergency situations like a pandemic, the parametrisation of mathematical models needs best be done as a continuous online data retrieval, in order to make future expert systems successful. The data retrieval aspect needed for such a strategy becomes better and better due to the development of IT infrastructures in different countries. There is another aspect related to Bayesian model selection. All selected models will only be the best Bayesian choice for some limited time, until the empirical data suggest new underlying mechanisms are at work which have not previously be incorporated. This can be interpreted as a kind of \emph{early warning system}, if properly implemented: there are new trends detected in the data suggesting underlying mechanisms are not all time invariant any more, some new process partially governs the pandemic situation.
\end{description}

\section*{Acknowledgements}	
MK likes to give special thanks to Odo Diekmann, Johannes M\"uller, Joan Saldana and Peter Wrede for reading parts of the manuscript and contributing references and ideas. \\

LMK acknowledges support from the European Union Horizon 2020 research and innovation programmes under the Marie Skłodowska-Curie grant agreement No. 777826 (NoMADS) and No. 691070 (CHiPS), the Cantab Capital Institute for the Mathematics of Information and Magda- lene College, Cambridge (Nevile Research Fellowship).

\bibliographystyle{siam.bst}
\bibliography{references}


\appendix

\section{Parameters}

Here we list the sets of parameters used throughout the simulation scenarios in this article. If a parameter does not appear in the list, it is equal to the parameter of the base model.
We like to emphasize that parameters are chosen according to rules of thumb, they should resemble reasonable values. However, none of these parameters has been derived from real world data. This parameter estimation problem we leave to future publications.
We also provide here the names of the jupyter notebooks used to create the according figures.

\begin{longtable}[c]{||>{\centering}m{0.1\linewidth} | m{0.35\linewidth} | m{0.4\linewidth} | m{0.12\linewidth} ||}
\caption{Parameters for simulations} \label{table:sec:simulation_parameters}
\endfirsthead
\endhead
	\hline
	\multicolumn{2}{|| l }{\textbf{model}} & \multicolumn{2}{r ||}{{\tt{jupyter notebook}}}\\\hline
	Parameter & Value & Meaning & Fig.\\ [0.5ex] 
	\hline\hline
	\multicolumn{2}{|| l }{\textbf{SIRD (base model)}} & \multicolumn{2}{r ||}{{\tt{RBSEa1\_SIRD.ipynb}}}\\\hline
    $N$ & $1000$ & total population & \multirow{7}{=}{\ref{fig:plotSIRD(SSA-ODE)}}\\ \cline{1-3}
    $I_0^0$ & $3$ & initial infectious & \\\cline{1-3}
    $T$ & $300$ & maximum time & \\\cline{1-3}
    $\iota^0$ & $0.2$ & infection rate & \\\cline{1-3}
    $\rho^0$ & $0.02$ & recovery rate & \\\cline{1-3}
    $\delta^0$ & $0.01$ & death rate & \\\cline{1-3}
    seed & $0$ & random seed & \\\hline
	\multicolumn{2}{|| l }{\textbf{SIRD$_\mathbf{a}$ (age)}} & \multicolumn{2}{r ||}{{\tt{RBSEa1\_SIRDa.ipynb}}}\\\hline
    $a \in A$ & \{1, 2, 3\} & index for age classes & \multirow{8}{=}{\ref{fig:plotAge(SSA)}}\\\cline{1-3}
    $I_0$ & $(1, 1, 1)$ & initial infectious & \\\cline{1-3}
    $p$ & $(0.25, 0.5, 0.25)$ & initial population shares & \\\cline{1-3}
    $C$ & $\left(\begin{smallmatrix} 76 & 38 & 5\\ 38 & 67 & 13\\ 5 & 13 & 9\end{smallmatrix}\right)$ & contact matrix & \\\cline{1-3}
    $\rho$ & $\rho^0 (1.5, 1.0, 0.5)$ & recovery rates & \\\cline{1-3}
    $c = q \cdot s \cdot i$ & $0.0047567053338882$ & product of calibration, infectivity and susceptibility factors & \\\cline{1-3}
    $\delta$ & $\delta^0 (0.25, 1.0, 2.0)$ & death rates & \\\hline
    $\iota$ & $c \cdot C$ & infection rates (${R_0 \approx 6.67}$) & \ref{fig:plotAge(SSA)_Contact}\\\hline
    $\iota$ & $c \cdot \left(\begin{smallmatrix} 1.0 & 1.0 & 0.2\\ 1.0 & 1.0 & 0.2\\ 0.2 & 0.2 & 1.0\end{smallmatrix}\right) \circ C$ & infection rates with reduced contact to elderly class (${R_0 \approx 6.59}$)\newline ($\circ$: Hadamard product) & \ref{fig:plotAge(SSA)_ContactRed}\\\hline
    \multicolumn{2}{|| l }{\textbf{SIRDresp (functional response)}} & \multicolumn{2}{r ||}{{\tt{RBSEa1\_SIRDresp.ipynb}}}\\\hline
    $T$ & $500$ & maximum time & \multirow{2}{=}{\ref{fig:plotSIRDres(SSA)}}\\\cline{1-3}
    $\iota(\nI; \mu)$ & $\frac{\iota^0}{1 + \mu \frac{\nI}{N}}$ & infection rate with functional response & \\\hline
    $\mu$  & $1$ & functional response parameter & \ref{fig:plotSIRDresp(SSA)_lambda1}\\\hline
    $\mu$  & $10$ & functional response parameter & \ref{fig:plotSIRDres(SSA)_lambda10}\\\hline
    \multicolumn{2}{|| l }{\textbf{SIRD$_c$resp (testing, functional response)}} & \multicolumn{2}{r ||}{{\tt{RBSEa1\_SIRDcresp.ipynb}}}\\\hline
    $T$ & $500$ & maximum time & \multirow{13}{=}{\ref{fig:plotSIRDtres(SSA)}}\\\cline{1-3}
    $c \in $ & $\{0, 1, 2\}$ & index for test classes & \\\cline{1-3}
    $I_0$ & $(3, 0, 0)$ & initial infectious & \\\cline{1-3}
    $\theta$ & $0.04$ & rate of testing& \\\cline{1-3}
    $\theta_S$ & $\theta$ & rate of testing of susceptibles & \\\cline{1-3}
    $\theta_I$ & $\theta$ & rate of testing of infected & \\\cline{1-3}
    $\theta_R$ & $\theta$ & rate of testing of recovered & \\\cline{1-3}
    $\lambda$  & $0.1$ & loss of information & \\\cline{1-3}
    $\iota$ & $\iota(\nI; \mu) \left(\begin{smallmatrix} 0.7 & 0.2 & 0.9\\ 0.9 & 0.2 & 1.\end{smallmatrix}\right)$ & infection rates & \\\cline{1-3}
    $\iota(\nI; \mu)$ & $\frac{\iota^0}{1 + \mu \frac{\nI_1}{N}}$ & infection rate with functional response dependent on tested infectious & \\\cline{1-3}
    $\rho$ & $\rho^0 (1., 1.5, 1.)$ & recovery rates & \\\cline{1-3}
    $\delta$ & $\delta^0 (1.0, 0.5, 1.0)$ & death rates & \\\hline
    $\mu$  & $1$ & functional response parameter & \ref{fig:plotSIRDtresp(SSA)_lambda1}\\\hline
    $\mu$  & $10$ & functional response parameter & \ref{fig:plotSIRDtres(SSA)_lambda10}\\\hline
  \end{longtable}

\end{document}